\documentclass[11pt,a4paper]{article}
\usepackage{jheppub}
\usepackage{float}
\usepackage{xcolor}
\usepackage{comment}
\usepackage{ulem}
\usepackage{mathrsfs}

\usepackage{amsmath,amssymb,hyperref,graphicx,amsthm,bm,subcaption,mathtools,appendix}
\usepackage{bbm}
\usepackage{tikz,tikz-feynman,tikz-cd}
\usepackage{environ}
\usepackage{breqn}
\usepackage{amsmath,amssymb,graphicx,amsthm,bm,mathtools,environ}

\NewEnviron{eqn}{
\begin{align}
\begin{split}
  \BODY
\end{split}
\end{align}
}
\NewEnviron{eqn*}{
\begin{align*}
\begin{split}
  \BODY
\end{split}
\end{align*}
}

\newcommand*{\ii}{\mathrm{i}}

\newcommand*{\mcS}{\mathcal{S}}

\newcommand{\g}{\gamma}

\newcommand{\gc}{\tilde{\gamma}}
\newcommand{\gb}{\hat{\gamma}}

\title{Carrollian ABJM: Fermions and Supersymmetry}

\date{}
\author[a]{Arjun Bagchi,} \author[b]{Arthur Lipstein,} \author[a]{Saikat Mondal,} \author[b]{and Alex Jiayi Zhang.} \author{\\}
\affiliation[a]{Indian Institute of Technology Kanpur, Kanpur 208016, India. \\}
\affiliation[b]{Department of Mathematical Sciences, Durham University, Stockton Road, DH1 3LE, Durham,
United Kingdom.\\}
\emailAdd{abagchi@iitk.ac.in}
\emailAdd{arthur.lipstein@durham.ac.uk}
\emailAdd{saikatmd@iitk.ac.in}
\emailAdd{jiayi.zhang@durham.ac.uk}
\abstract{A natural approach for constructing a concrete example of flat space holography is to take the flat space limit of a well-understood example of AdS/CFT, such as the one relating M-theory in AdS$_4$ times an orbifolded 7-sphere to a certain three dimensional superconformal Chern-Simons-matter theory known as the ABJM theory living in the boundary of AdS$_4$. In particular, taking the flat space limit of the bulk corresponds to taking the speed of light $c$ to zero in the boundary, giving rise to a Carrollian superconformal theory. This limit is subtle to implement for fermions, however, since the Dirac algebra is sensitive to the spacetime metric and therefore takes a different form in Carrollian spacetimes than it does in Minkowski space. In fact, we show that there are four possible ways of realising Carrollian fermions, one of which arises at leading order in the $c\rightarrow0$ limit of relativistic fermions. In three dimensions, there is an additional complication that the minimal realisation of the Carrollian Dirac algebra requires $4\times4$ matrices rather than $2\times 2$ matrices familiar from the relativistic case. Nevertheless, we show that the $c\rightarrow0$ limit of the ABJM theory can be recast in terms of Carrollian Dirac matrices and enjoys an infinite-dimensional Carrollian superconformal symmetry whose bosonic subsector is the extended BMS$_4$ algebra encoding the asymptotic symmetries of four dimensional Minkowski space. This provides a concrete starting point for constructing a Carrollian gauge theory dual to M-theory in flat space.}

\begin{document}

\maketitle

\section{Introduction}
The holographic principle \cite{tHooft:1993dmi, Susskind:1994vu} remains our most hopeful tool for the construction of a theory of quantum gravity. In stark contrast to the overwhelming success of the AdS/CFT correspondence \cite{Maldacena:1997re}, holographic formulations in asymptotically de Sitter and asymptotically flat spacetimes remain less successful and less explored. 

In recent years, there has been a renewed effort in formulating flatspace holography. There are two principal avenues in these recent explorations, the so-called Celestial and Carrollian approaches. Celestial holography \cite{Strominger:2017zoo, Pasterski:2021raf, Raclariu:2021zjz} is built out of the observation that the Lorentz group in four dimensions acts as the 2d global conformal group at the asymptotic boundary of flat space and aims to build a 2d hologram in terms of a relativistic 2d CFT living on the celestial sphere.

Carrollian holography \cite{Bagchi:2010eg, Bagchi:2012cy, Bagchi:2016bcd, Bagchi:2025vri, Nguyen:2025zhg, Ruzziconi:2026bix} on the other hand puts forward a co-dimension one putative dual theory, where the field theory is a Carrollian CFT living on the entire null boundary. A Carrollian CFT can be obtained from a relativistic CFT by sending the speed of light $c$ to zero and is naturally defined on a null surface. A co-dimension one approach is also closer in spirit to the original AdS/CFT correspondence and indeed one can get to Carrollian holography by a suitable $R_{AdS} \to \infty$ limit of AdS/CFT \cite{Bagchi:2023fbj, Bagchi:2023cen, Alday:2024yyj, Lipstein:2025jfj}. $R_{AdS}^{-1}$ in this formulation can be viewed as an effective speed of light in the boundary theory, so sending $R_{AdS} \to \infty$ in the bulk to reach flat space is equivalent to sending $c\to0$ on the boundary to reach a Carrollian CFT \cite{Bagchi:2012cy, Bagchi:2016bcd}. 

Both Celestial and Carrollian holography have been principally bottom-up approaches, and have relied on symmetries to build the respective holograms. While Celestial holography has had success in uncovering new directions in asymptotic symmetries and scattering amplitudes, building a bonafide Celestial CFT from scratch remains a challenge (see for example \cite{deBoer:2003vf,Ball:2019atb,Costello:2023hmi} for progress along this direction). On the other hand, one can write down explicit Carrollian CFTs, like the free scalar theory, which have some of the desired properties of a hologram of 4d asymptotically flat space (AFS), like distributional correlation functions \cite{Bagchi:2022emh}. The fact that Carrollian CFTs are obtainable from a $c\to0$ limit of relativistic CFTs also offer the significant advantage of starting with any consistent relativistic 3d CFT and building a 3d Carroll CFT out of it. 

One of the main goals of the flatspace holography programme is of course to find a concrete flatspace hologram. We would be interested especially to find a dual to 4d AFS, since at distances below cosmological scales, this provides a good approximation of our Universe. In this paper, we take concrete steps to building a top-down model of a flatspace hologram. Our tool will be the AdS$_4$/CFT$_3$ correspondence, which relates M-theory in AdS$_4 \times$S$^7/\mathbb{Z}_{k}$ to a superconformal Chern-Simons theory known as the ABJM theory \cite{Aharony:2008ug}. In \cite{Bagchi:2024efs}, a toy model consisting of Chern-Simons fields coupled to scalars was considered and shown to exhibit infinite-dimensional Carrollian conformal symmetry. The goal of this paper will be to incorporate fermions and supersymmetry into this story.

Fermions in the Carroll world are intricate objects \cite{Bagchi:2022eui}. Due to the degenerate nature of null manifolds, there are zeros in the metric defined on them. This means the resulting Clifford algebra for null or Carroll manifolds inherits these degeneracies. As we will elaborate in the later sections, the degenerate Carroll Clifford algebra leads to different notions of fermions. We will be motivated by the $c\rightarrow 0$ limit of relativistic fermion theories and will first consider the simplest type of fermions in this work. In a companion paper \cite{Bagchi:2026lgk}, we work out the Carroll expansion of fermions, where instead of sending the speed of light ($c\to0$) to zero, we do an order by order expansion around $c=0$, following earlier constructions for bosonic fields in \cite{deBoer:2021jej}. In this paper, we will work exclusively with the leading order theory, which is expected to capture the scattering amplitudes of the putative bulk theory through Modified Mellin transformations \cite{Banerjee:2018gce, Bagchi:2022emh}, although subleading terms may need to be included as well prior to taking the $c\rightarrow 0$ limit for reasons we explain in section \ref{props}. One of the subtle details is the understanding of how degenerate Carrollian gamma matrices combine rather beautifully to reproduce answers which one naively expects from the $c\to0$ limit. We will then demonstrate how Carrollian supersymmetry works and show how to build a Carroll version of ABJM. The Carrollian ABJM Lagrangian so constructed will be shown to exhibit superconformal Carroll symmetries \cite{Bagchi:2022owq, Lipstein:2025jfj} that are expected from a theory dual to supergravity in flat spacetime. 

One of the remarkable aspects of holography in AFS is that the asymptotic symmetry group in AFS is infinite dimensional in all spacetime dimensions. In particular, in 4d AFS, the bosonic symmetries of interest are the so-called 4d Bondi-van der Burgh-Misner-Sachs (BMS$_4$) algebra. In the presence of supersymmetry, this enhances to the Super BMS$_4$, which is also infinite dimensional. On the other hand, the relativistic ABJM theory exhibits a finite dimensional relativistic superconformal symmetry. We will show in what follows how the Carroll version of ABJM naturally encodes this infinite dimensional enhancement of symmetries. This work sets the stage for a thorough investigation of the Carrollian ABJM model and is the first step to building of a novel top-down flatspace hologram derived from the AdS$_4$/CFT$_3$ correspondence.

This paper is organised as follows. In section \ref{carrollrev} we review Carrollian conformal symmetry. After that, we describe how to construct Carroll fermion theories in section \ref{carrollfermi}, where we also provide a toy model of Carrollian supersymmetry. In section \ref{abjmcar} we then review the ABJM theory and explain how to take its $c\rightarrow 0$ limit, recasting the resulting action in terms Carrollian Dirac matrices. After that we show that the theory enjoys an infinite dimensional Carrollian superconformal symmetry and spell out the symmetry algebra in section \ref{abjmsymmetry}. Finally, we present our conclusions and future directions in section \ref{conclusion}. There are also a number of Appendices which provide further technical details.  

\section{Carroll and Conformal Carroll Symmetries} \label{carrollrev}

In this section we review some basics of Carrollian symmetry to set the stage for the paper. 

\subsection{Carrollian symmetries}
We begin by briefly recalling the basic structures of Carrollian manifolds. A Carroll structure is defined as the triplet $(\mathcal{M},h_{\mu\nu}, \theta^\rho)$ where $\mathcal{M}$ is $d$-dimensional manifold, $h_{\mu\nu}$ a symmetric covariant tensor  of rank $(d-1)$ and signature $(0, + \ldots +)$ and $\theta^\mu$ is a vector field that generates its kernel, i.e. $h_{\mu\nu} \theta^\nu = 0.$ A flat Carroll manifold is characterised by 
\begin{eqnarray}\label{FlC}
h_{\mu\nu} = \begin{pmatrix}
			0 & 0 \\
			0 & \quad I_{d-1}
		\end{pmatrix}, 
		\qquad
\Theta^{\mu\nu} = 	\begin{pmatrix}
			1 & 0 \\
			0 & \quad 0_{d-1} 
		\end{pmatrix}   \, = \theta^\mu \theta^\nu.
\end{eqnarray}      
The Lie algebra of isometries of this flat Carroll structure gives rise to 
Carroll algebra, the non-vanishing commutators of which are as follows:
\begin{align}\label{carral}
        [J_{ij},J_{kl}]&=so(d), \quad [C_i,P_j]=-\delta_{ij}H, \quad [J_{ij},X_{k}]=-\delta_{ik}X_{j} + \delta_{jk}X_{i}.
\end{align}     
In the $d$-dimensional manifold where one has a coordinate chart $\{t, x^1\ldots x^{d-1}\}$, the generators $J_{ij} = x_{[i}\partial_{j]}$ are the rotations, $X_{i}\in\left\{ P_{i} = \partial_i,\, C_{i}= x_i\partial_t \right\} $ are the momenta and Carroll boost generators, and $H= \partial_t$ is the Hamiltonian. Notice that the Carroll boosts commute and the Carroll Hamiltonian is a central element of the algebra. The above algebra can be derived from an In{\"o}n{\"u}-Wigner contraction of the Poincaré algebra where the speed of light $c$ is dialled to zero. In addition to the above finite generators, the isometry algebra interestingly allows for spatial dependent translations of the null time direction
\begin{align}\label{st}
    M(f) = f(x^i) \partial_t. 
\end{align}
These are called supertranslations and reduce to just the Hamiltonian and Carroll boosts when $f(x)$ is truncated to linear order in $x$. 

\subsection{Conformal Carroll and BMS} \label{bosonicsymm}
The conformal isometries of the flat Carroll structure discussed above 
\begin{align}
    \pounds_{\xi}h_{\mu\nu}= 2\lambda h_{\mu\nu} , ~~~\quad \pounds_{\xi}\theta^{\mu}= - \lambda \theta^{\mu} \,
\end{align}
generate the conformal Carroll algebra. The conformal Killing vector field is of the form
\begin{align}\label{CC-killing}
   \xi= f(x^k)\partial_t + \omega^i_{\,j}\,x^j\partial_i + b^i\partial_i + \Delta\left(t\partial_t + x^i\partial_i\right) + \kappa_i\left(2x^i(t\partial_t + x^j\partial_j)-x^k x_k\delta^{ij}\partial_j\right)\,,
\end{align}
where  $\omega^i_{\, j}$, $b^i$, $\Delta$ and $\kappa_i$ are coefficients for rotations, translations, dilatation and spatial special conformal transformations respectively, while the first term again represents the infinite dimensional supertranslations. These close into the infinite-dimensional conformal Carroll algebra (CCA). The CCA in $d$-dimensions is isomorphic \cite{Bagchi:2010eg, Duval:2014uva} to the Bondi-van der Burgh-Metzner-Sachs (BMS) algebra \cite{Bondi:1962px,Sachs:1962zza} in $(d+1)$ dimensions, the asymptotic symmetry algebra of asymptotically flat spacetime in $(d+1)$ dimensions at its null boundary: 
\begin{align}
\mathfrak{cca}_{d} = \mathfrak{bms}_{d+1}. 
\end{align}
This is very similar to a central point of the AdS/CFT correspondence where the isometries of AdS are realised as the conformal symmetries of the boundary field theory. In this case, the (asymptotic) isometries of flatspace are mapped to the infinite dimensional conformal symmetries of the field theory living on the null boundary (which is Carrollian be definition). This is the first step in the formulation of holography in asymptotically flat spacetimes in terms of a co-dimension one dual field theory governed by the conformal Carroll algebra \cite{Bagchi:2010eg, Bagchi:2016bcd}. Initial successes of this correspondence was in three bulk and two boundary dimensions \cite{Bagchi:2012cy, Barnich:2010eb, Barnich:2012aw, Bagchi:2012yk, Bagchi:2012xr, Barnich:2012xq, Bagchi:2014iea, Bagchi:2015wna, Jiang:2017ecm, Hartong:2015usd} and the recent resurgence of correspondence between 4d bulk and 3d boundary theories has followed from \cite{Donnay:2022aba, Bagchi:2022emh}. This is the case we will also be interested in and now we turn our focus here. 

\medskip

In $d=3$ boundary dimensions, there is a further enhancement of the conformal isometries of Carroll manifolds beyond the above mentioned super-translations to include so-called super-rotations. The CCA$_3$ or equivalently the BMS$_4$ algebra closes to form 
\begin{subequations}\label{eq:BMS_alg}
\begin{align}
    &[L_n,L_m] = (n-m)L_{n+m}\,,\quad [\bar{L}_n,\bar{L}_m] = (n-m)\bar{L}_{n+m}, \quad [M_{r,s},M_{p,q}] = 0, \\
    &[L_n, M_{r,s}] = \left(\frac{n+1}{2}-r\right)M_{n+r,s}\,,\quad [\bar{L}_n, M_{r,s}] = \left(\frac{n+1}{2}-s\right)M_{r,s+n} \, . 
\end{align}
\end{subequations}
In terms of the geometry at $\mathscr{I}^\pm$, which is $\mathbb{R}_t\times$S$^2$,  $M_{rs}$ are the modes of the supertranslations encountered above \eqref{st}, i.e. translations of the null direction $t$ which depend on spatial coordinates $z, \bar{z}$, the stereographic coordinates of the sphere. $L_n, \bar{L}_n$ are superrotations that  generalise Lorentz generators to 2d conformal generators on the celestial sphere. 
A particularly useful differential representation of the conformal Carroll generators is
\begin{align}\label{eq:BMS4_diff_op}
    L_n = z^{n+1}\partial_z + \frac{1}{2}(n+1)z^nt\partial_t\,,\quad \bar{L}_n = \bar{z}^{n+1}\partial_{\bar{z}} + \frac{1}{2}(n+1)\bar{z}^nt\partial_t\,,\quad M_{r,s} = z^r\bar{z}^s\partial_t\,,
\end{align}
where $z=x+iy$ and $\bar{z} = x-iy$ are stereographic coordinates on the celestial sphere. The subalgebra $n,m = \{-1,0,1\}$ and $r,s = \{0,1\}$ describes a 4d Poincare subalgebra. It is convenient to define the following relations: 
\begin{subequations}\label{eq:BMS4_global_identification}
\begin{align}
    &L_{-1} = \frac{1}{2}(P_x - iP_y)\,,\quad L_0 = \frac{1}{2}(D+iJ_{xy})\,,\quad L_{+1}= \frac{1}{2}(K_x + iK_y) \, , \\
    &\bar{L}_{-1}= \frac{1}{2}(P_x + iP_y)\,,\quad \bar{L}_0 = \frac{1}{2}(D-iJ_{xy})\,,\quad \bar{L}_{+1}= \frac{1}{2}(K_x - iK_y)\, ,\\
    &M_{0,0} = P_0\,,\quad M_{0,1} = C_x - iC_y\,,\quad M_{1,0} = C_x+iC_y \,,\quad M_{1,1} = K_0\, .
\end{align}
\end{subequations}
This is now a map between the 4d Poincaré algebra and the 3d Conformal Carroll algebra which is the algebra obtained by taking $c\to 0$ limit of the 3d relativistic conformal algebra. The generators $\left\{J_{ij}, P_{i},C_{i}, H\right\}$ generate the Carroll algebra in \eqref{carral}. In addition, we have dilatation $D$ and special conformal generators $K_0$, $K_i$. The additional non-vanishing brackets are given by 
\begin{align}\label{confc}
    &[D,P_i] = P_i\,, \,~~~ [D,H] = H\,, \,~~~ [D,K_i] = -K_i\,, \,~~~ [D,K_0] = -K_0, \nonumber\\
&[K_0,P_i] = -2C_i\,, \,~~~ [K_i,H]=-2C_i\,, \,~~~ [K_i,P_j] = -2\delta_{ij}D-2J_{ij}\, \nonumber\\
&[K_i, C_j] = \delta_{ij}K_0\,,\quad [J_{ij},K_l] = \delta_{il}K_j - \delta_{jl}K_i.
\end{align}
Together with the Carroll algebra, these relations define the global conformal Carroll algebra. This is equivalently obtained from \eqref{CC-killing} by setting $d=3$ and taking the arbitrary function $f(x^k)$ to be linear, as was done for the Carroll algebra earlier. 

\subsection{Carroll CFTs: representations} \label{carrolcftrep}
We will be interested in quantum field theories which are governed by the CCA$_3$. In these theories, we will label operators $\mathcal{O}$ by their weights under $L_0, \bar{L}_0$: 
\begin{align}
[L_0,\mathcal{O}(0)] = h \mathcal{O}(0) = \frac{1}{2} (\Delta-\ii J) \mathcal{O}(0), \quad [\bar{L}_0, \mathcal{O}(0)] = \bar{h} \mathcal{O}(0) = \frac{1}{2}(\Delta+\ii J) \mathcal{O}(0).
\end{align}
We would also like to work with highest weight representations, where Carrollian primary operators are defined as:
\begin{align}
[L_n,\mathcal{O}(0)] = 0, \quad [\bar{L}_n, \mathcal{O}(0)] = 0 \quad \forall n>0, \qquad [M_{r,s}, \mathcal{O}(0)] = 0 \quad \forall \, r>0 \, \text{and} \, s>0. 
\end{align}
For a primary highest weight operator $\mathcal{O} $ at a generic point in the 3d Carroll manifold labelled by $(t, z, \bar{z})$, the action of the 3d CCA generators is given by \cite{carroll_csm}
\begin{eqn}\label{eq:BMS_general}
[L_n,\mathcal{O}(t, z, \bar{z})] &= \Bigl[z^{n+1}\partial_z +\frac{1}{2}z^n(n+1)(\Delta-\ii J +t\partial_t   )  \Bigr] \mathcal{O}(t, z, \bar{z}) \\
[\bar{L}_n, \mathcal{O}(t, z, \bar{z})] &= \Bigl[\bar{z}^{n+1}\partial_{\bar{z}} +\frac{1}{2}\bar{z}^n(n+1)(\Delta + \ii J  +t\partial_t   )  \Bigr] \mathcal{O} (t, z, \bar{z})\\
[M_{mn},\mathcal{O}(t, z, \bar{z})] &= z^m\bar{z}^n\partial_t \mathcal{O}(t, z, \bar{z}), \\
\end{eqn}
where $J$ denotes the (anti-hermitian) generator of internal spatial rotations.

\section{Carroll Fermions} \label{carrollfermi}

The definition of Dirac fermions on Carrollian manifolds is not unique due to the degenerate nature of the metric. In this section we will spell out four different formulations of Carollian fermions and show that one arises at leading order in the $c \rightarrow 0$ limit of relativistic Dirac fermions. We also describe additional subtleties that arise in odd dimensions and a toy model of Carrollian supersymmetry. 

\subsection{Basic features}
In this subsection, we quickly summarize the basic features of Carrollian fermions \cite{Bagchi:2022eui, Bagchi:2026lgk, Grumiller:2025rtm}. Due to the degenerate structure of the Carrollian manifold, the Clifford algebra is modified to its Carrollian version:
\begin{align}\label{CClf} 
\{\gc_{\mu},\gc_{\nu}\}=2h_{\mu\nu},~~\quad \{\gb^{\mu},\gb^{\nu}\}=2\Theta^{\mu\nu}\,.
\end{align}
This naturally leads to two inequivalent classes of fermions, which we refer to as ``lower'' ($\downarrow$) and ``upper'' ($\uparrow$) fermions corresponding to the type of gamma matrices actions are built out of. The degenerate nature of the Clifford algebra has several interesting consequences. We sequentially list these below. 

\begin{figure}[t]
    \centering
    \includegraphics[width=1\textwidth]{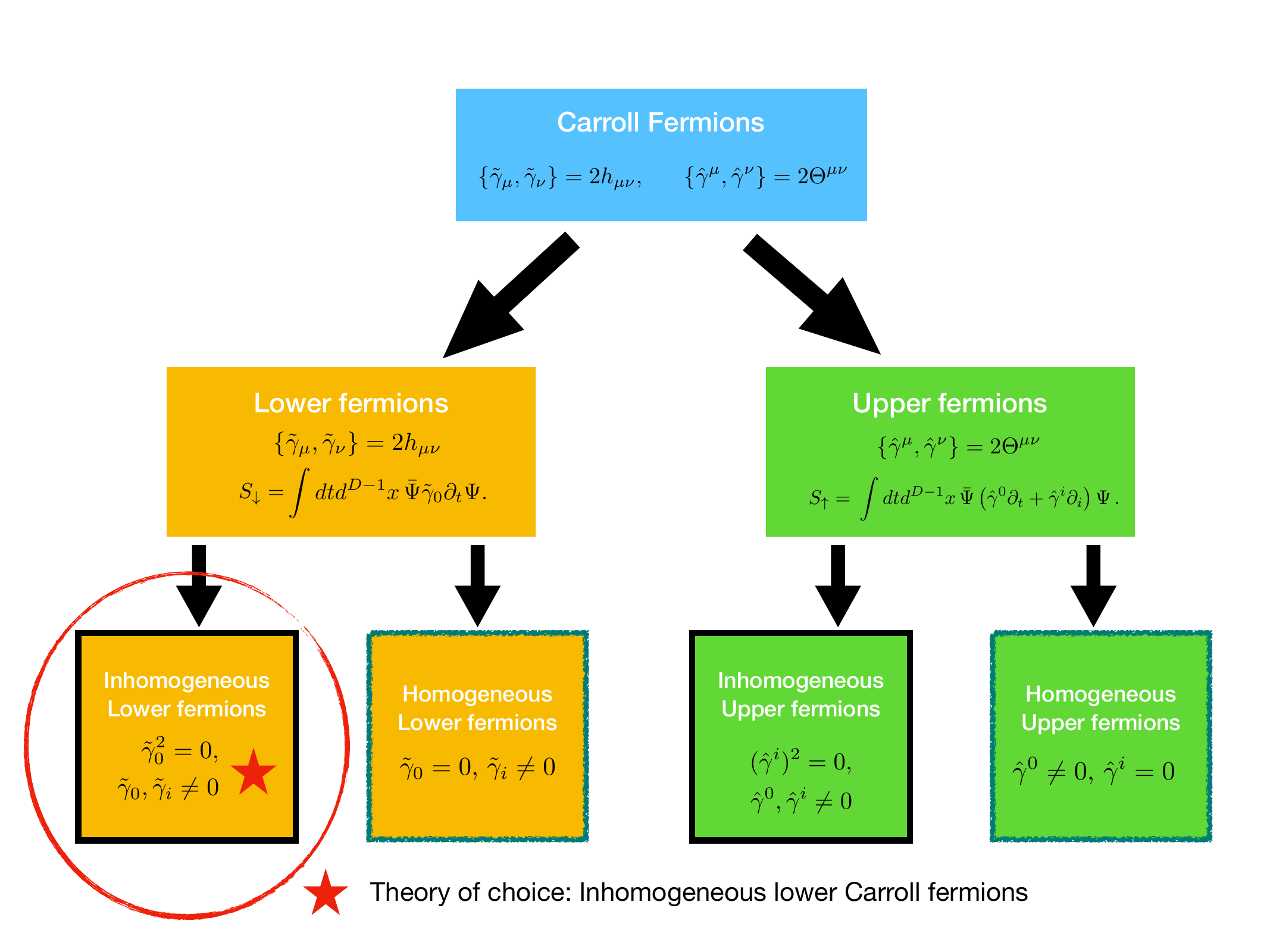}
    \caption{Summary of Carroll fermions}
    \label{fig1}
\end{figure}

\begin{itemize}

    \item{\em Spacetime symmetries}: As a consistency requirement, the spacetime algebra obtained from the Carroll Clifford algebra should close to the Carroll algebra \eqref{carral} restricted to rotations and boosts. This is indeed the case. We define\footnote{Here the superscripts $\downarrow$ and $\uparrow$ are introdcued solely for notational distinguishabilty. We emphasize that the symbol $\downarrow$ appears exclusively on matrices with lower indices.}
    \begin{align}\label{sig}
        \Sigma^\downarrow_{\mu\nu} = \frac{1}{4} [\gc_\mu, \gc_\nu].
    \end{align}
    It can be easily shown that ${\Sigma^{\downarrow}}$ generate the Carroll algebra \eqref{carral}. The case of the upper gammas is a bit more subtle and we refer the reader to \cite{Bagchi:2022eui} for more details. 
    
    \item {\em Adjoint}: A feature of the Carroll spinors that is distinct from their relativistic counterparts is the definition of the adjoint. Defining the adjoint via $\gc_0$ needs to change as this is not Carroll invariant and instead is given by $\bar{\Psi}= \Psi^{\dagger}\Lambda$, where $\Lambda$ is a Hermitian matrix to be fixed by symmetry requirements. The generators are required to satisfy ${\Sigma^{\downarrow, \uparrow}}^\dagger = -\Lambda\Sigma^{\downarrow, \uparrow}\Lambda^{-1}$. 

   \item{\em Spinor transformations}: Under Carroll transformations, spinor $\Psi(x)$ transforms as 
	\begin{align}
		\Psi(x) \rightarrow \mathcal{S}^{\downarrow, \uparrow}[\Sigma]\Psi(x),
	\end{align}
where $\mathcal{S}^{\downarrow, \uparrow}[\Sigma] = \exp\big(\frac{1}{2}\omega \cdot \Sigma^{\downarrow, \uparrow} \big).$
Here $\Sigma$'s are defined in \eqref{sig} and $\omega \equiv \omega^{\mu\nu}$'s are antisymmetric transformation parameters. When constructing ${\downarrow, \uparrow}$ fermion theory, the indices of the Carroll generators $\Sigma$ are chosen consistently as lower or upper respectively. 

    \item {\em Actions}: Two sets of fermions $(\downarrow, \uparrow)$ lead to different actions. For $\downarrow$ fermions, we have:  
\begin{align}
    S_{\downarrow} = \int d^dx \,\bar{\Psi}\,\theta^{\mu}\theta^{\nu}\,\tilde{\gamma}_{\mu}\partial_{\nu}\, \Psi = \int dtd^{d-1}x \,\bar{\Psi}\gc_{0}\partial_{t}\Psi . 
    \label{actionleading}
\end{align}
These $\downarrow$ fermions have a distinct Carrollian flavour as the action only contains a time derivative and spatial derivatives vanish. For $\uparrow$ fermions, we instead get: 
\begin{align}
    S_{\uparrow} = \int d^dx \bar{\Psi}\gb^{\mu}\partial_{\mu}\Psi = \int dtd^{d-1}x\, \bar{\Psi}\left(\gb^0\partial_t + \gb^i\partial_i\right)\Psi\,.
    \label{actionsubleading}
\end{align}
These $\uparrow$ fermions are more relativistic in nature since both derivatives appear. The two actions can be obtained as leading (corresponding to $\downarrow$) and subleading (corresponding to $\uparrow$) pieces of the relativistic fermionic action by appropriately expanding in the limit where the speed of light vanishes. For details of this, the reader is pointed to the companion paper \cite{Bagchi:2026lgk}. In modern parlance, the $\downarrow$ fermions can be called electric fermins while the $\uparrow$ fermions are magnetic. In this paper, we will focus entirely on the leading $\downarrow$ electric fermions. 

    \item {\em Representations}: Carroll Clifford algebras are degenerate and contain nilpotent matrices. By the rank-nullity theorem, a $n\times n$ nilpotent matrix has rank $\leq n-1$. Hence, $2\times 2$ nilpotent $\gamma$'s have either rank $0$ or rank $1$, giving rise to two inequivalent representations \cite{Bagchi:2026lgk}. The rank $0$ one is called the \textit{homogeneous} representation $\mathcal{R}_H$, whereas the rank $1$ is called \textit{inhomogeneous} representation $\mathcal{R}_I$. For $\mathcal{R}_H$, the nilpotent matrix is trivially zero, while the other gamma matrices spanning the Clifford algebra can be chosen among the Pauli spin matrices. In higher-dimensional cases, $\mathcal{R}_H$ follows this same pattern, whereas $\mathcal{R}_I$ can be systematically built from the lower-dimensional gamma matrices. 
\end{itemize}
A schematic overview of different Carroll fermions is presented in the figure \ref{fig1}. We will choose to work with the Inhomogeneous Lower fermions in what follows, which we denote by the representation $\mathcal{R}^{\downarrow}_I$. 

\medskip

The obvious question is why we choose this one out of the a priori four choices we seem to have. First let's decide on lower versus upper fermions. Our focus in the second half of the paper would be in constructing a putative dual of 4d asymptotically flat spacetime from the Carroll limit of a certain superconformal Chern-Simons theory dual to M-theory in AdS$_4 \times$S$^7/\mathbb{Z}_k$, known as the ABJM theory. We will focus on the 
leading terms in the $c\to0$ limit. As stated above and shown in \cite{Bagchi:2026lgk}, the lower fermions are the ones which arise at leading order in this expansion. As to the choice between Homogeneous and Inhomogeneous, the answer is simple. As can readily be seen from \eqref{actionleading}, for the Homogeneous choice, the action trivially vanishes: 
\begin{align}
   \mathcal{R}^{\downarrow}_H:  \gc_0 =0 \Rightarrow S_{\downarrow} = 0.
\end{align}
So lower fermions with the Homogeneous choice of gamma matrices would lead to a trivially vanishing action. We thus focus on $\mathcal{R}^{\downarrow}_I$. 

\medskip

For even spacetime dimensions, the gamma matrices in $\mathcal{R}^{\downarrow}_I$,  are of the following form: 
\begin{subequations}
    \begin{align} \text{2D:} \qquad
    \gc_0 &= \begin{pmatrix}
        0 & 0\\
        \ii & 0
    \end{pmatrix}\, \, \text{or} \begin{pmatrix}
        0 & \ii\\
        0 & 0
    \end{pmatrix}\,,  \quad \gc_1 = \begin{pmatrix}
        1 & 0\\
        0 & -1
    \end{pmatrix}\,,\quad \Lambda = \begin{pmatrix}
        0 & 1\\
        1 & 0
    \end{pmatrix}\\
 \text{4D:} \qquad 
    \gc_0 &= \begin{pmatrix}
        \textbf{0} & \textbf{0}\\
        i\textbf{I} & \textbf{0}
    \end{pmatrix}\, \text{or} \, \begin{pmatrix}
        \textbf{0} & \ii\textbf{I}\\
        \textbf{0} & \textbf{0}
    \end{pmatrix}\,,\quad \gc_i = \begin{pmatrix}
        \sigma^i & \textbf{0}\\
        \textbf{0} & -\sigma^i
    \end{pmatrix},\quad \Lambda = \begin{pmatrix}
        \textbf{0} & \textbf{I}\\
        \textbf{I} & \textbf{0}
    \end{pmatrix}
    \label{evendiracmatrices}
\end{align}
\end{subequations}

Specialising to 4D, we now reconsider the action in $\mathcal{R}^{\downarrow}_I$, written in terms of two-components spinors:
\begin{align}
\label{sdown1}
    S_{\downarrow}^{(1)} =  \int d^4x \,{\Psi}^\dagger \Lambda \,\Theta^{\mu\nu}\,\tilde{\gamma}_{\mu}\partial_{\nu}\, \Psi =  \int d^4x \,\begin{pmatrix}
        \psi^{\dagger}_L & \psi^{\dagger}_R 
    \end{pmatrix}\,\begin{pmatrix}
        \textbf{0} & \textbf{I}\\
        \textbf{I} & \textbf{0}
    \end{pmatrix}\, \begin{pmatrix}
        \textbf{0} & \textbf{0}\\
        \ii\textbf{I} & \textbf{0}
    \end{pmatrix}\, \partial_t\begin{pmatrix}
        \psi_L \\ \psi_R 
    \end{pmatrix}\,
    = \ii\int d^4x \, \psi^{\dagger}_L \partial_t \psi_L 
\end{align}
We see that the Carroll theory loses all information of the $\psi_R$ component and hence loses half its degrees of freedom compared to the relativistic theory. If we had instead chosen the other $\tilde{\gamma}_0$, we would have gotten
\begin{align}
\label{sdown2}
S_{\downarrow}^{(2)} = \int d^4x \,{\chi}^\dagger \Lambda \,\Theta^{\mu\nu}\,\tilde{\gamma}_{\mu}\partial_{\nu}\, \chi = \int d^4x \,\begin{pmatrix}
        \chi_L^\dagger & \chi_R^\dagger 
    \end{pmatrix}\,\begin{pmatrix}
        \textbf{0} & \textbf{I}\\
        \textbf{I} & \textbf{0}
    \end{pmatrix}\, \begin{pmatrix}
        \textbf{0} & \ii\textbf{I}\\
        \textbf{0} & \textbf{0}
    \end{pmatrix}\, \partial_t\begin{pmatrix}
        \chi_L \\ \chi_R 
    \end{pmatrix}\,
    = \ii\int d^4x \, \chi_R^\dagger \partial_t \chi_R 
\end{align}
The action of the spacetime generators on the spinors is given by 
\begin{align}
    \Psi^{\prime\downarrow}(x)= \exp\left(\frac{1}{2}\omega^{ab} \, \Sigma^{\downarrow}_{ab} \right)\Psi(x). 
\end{align}
In more detail, for the choice of $\gc_0$ in \eqref{sdown1} we obtain 
\begin{align}
    \begin{pmatrix} \psi'_L \\ \psi'_R \end{pmatrix}\, = (\textbf{I} + \omega^{0i} \Sigma^{\downarrow}_{0i}) \begin{pmatrix} \psi_L \\ \psi_R \end{pmatrix}\, = \begin{pmatrix} \psi_L \\ \psi_R + \omega^{0i} \sigma_i \psi_L \end{pmatrix}\
    \label{carrollsboost}
\end{align}
Notice that $\psi_L$, which appears in the action in \eqref{sdown1}, does not change under Carroll boosts. Hence, the action is trivially invariant under Carroll boosts since
\begin{align}
    \delta_C \psi_L = 0. 
\end{align}
Similarly, the action in \eqref{sdown2} associated with the other choice of $\gc_0$ is invariant under Carroll boosts since
\begin{align}
    \delta_C \, \chi_R = 0. 
\end{align}

\subsection{Limit from relativistic fermions: a conundrum ... and a solution.}

In a recent study of Carroll fermions in \cite{Bergshoeff:2023vfd} (see also \cite{Koutrolikos:2023evq}), the authors worked on a formulation where they kept the relativistic gamma matrices and work out actions which turn out to be identical to the two actions we have derived earlier $S_{\downarrow,\uparrow}$. Very clearly there is something amiss here. The relativistic gamma matrices would give rise to $\Sigma$ matrices which are relativistic and hence generate the Lorentz algebra instead of the Carroll algebra. Hence the spacetime symmetries of these fermions are clearly not Carrollian and in particular, the boost matrices constructed out of these gamma matrices don't commute. On the other hand, the fact that their actions match with ours indicates that by some miracle, things do work out even for the relativistic gamma matrices. So what is going on and how can we reconcile these apparent contradictions? 

\medskip

Given our discussion about $S_{\downarrow}^{(1)}$ and $S_{\downarrow}^{(2)}$, the solution is actually staring us in the face. We now consider an action
\begin{align}
    S = S_{\downarrow}^{(1)} + S_{\downarrow}^{(2)} & = \int d^4x \, i(\psi_L^\dagger \partial_t \psi_L + \chi_R^\dagger \partial_t \chi_R) \nonumber\\
    &= \int d^4x \,\begin{pmatrix}
        \psi_L^\dagger & \chi_R^\dagger 
    \end{pmatrix}\,\begin{pmatrix}
        \textbf{0} & \textbf{I}\\
        \textbf{I} & \textbf{0}
    \end{pmatrix}\, \left[\begin{pmatrix}
        \textbf{0} & \textbf{0}\\
        \ii\textbf{I} & \textbf{0}
    \end{pmatrix}\, + \begin{pmatrix}
        \textbf{0} & \ii\textbf{I}\\
        \textbf{0} & \textbf{0}
    \end{pmatrix}\,\right] \partial_t\begin{pmatrix}
        \psi_L \\ \chi_R 
    \end{pmatrix}\, \nonumber\\
    &= \int d^4x \, \bar{\Psi} \, \Gamma_0 \, \partial_t \Psi, 
\end{align}
where $\Psi$ and $\Gamma_0$ is a relativistic 4d Dirac spinor and Dirac matrix, respectively: 
\begin{align}
   \Psi = \begin{pmatrix} \psi_L \\ \chi_R \end{pmatrix},\,\,\,\Gamma_0 = \begin{pmatrix}
        \textbf{0} & \ii\textbf{I}\\
        \ii\textbf{I} & \textbf{0}
\end{pmatrix}\, ,
\end{align}
and $\bar{\Psi} =-\ii \Psi^\dagger (\Gamma_0)$ 
is the usual relativistic Dirac conjugate. This provides a simple way to cure the apparent contradictions we faced in the paragraph above. Given that a Carroll fermion loses only of its two Weyl components to leading order, this reconstruction is also natural.

One could complain that we don't have a first principles way of deriving why the relativistic object needed to be constructed out of two Carroll fermions with different choices of Carroll $\gc_0$'s. This becomes more aparent when one considers fermions in the lightcone. It has been recently shown that in lightcone coordinates, the Poincare algebra has two codimensional one Carroll subalgebras \cite{Bagchi:2024epw, Majumdar:2024rxg}. The relativistic Clifford algebra also naturally splits into two lower dimensional Carroll Clifford algebras, relating Carroll fermions to their higher dimensional relativistic cousins. The degrees of freedom of the relativistic theory can be shown to rearrange themselves on the two different null directions. The two different $\gc_0$'s correspond to $\gc_+$ and $\gc_-$ of the two lightcones. For further details on this rather beautiful structure, we point the reader to \cite{Bagchi:2026lgk}.

\subsection{What happens in odd dimensions?} 
The representation theory of the Carroll Clifford algebra exhibits additional subtleties in odd dimensions. Since the remainder of this paper focuses on three dimensions, we will consider this case in detail now. In three-dimensional Carrollian spacetime, the Clifford algebra is generated by ${\tilde{\gamma}_0,\tilde{\gamma}_1,\tilde{\gamma}_2}$ satisfying
\begin{equation}\label{cr3}
\{\tilde{\gamma}_0,\tilde{\gamma}_0\}=0, \qquad
\{\tilde{\gamma}_0,\tilde{\gamma}_i\}=0, \qquad
\{\tilde{\gamma}_i,\tilde{\gamma}_j\}=2\delta_{ij}, \quad i,j=1,2.
\end{equation}
The spatial gamma matrices generate $\mathrm{Cl}(2)$, whose irreducible complex representation is two-dimensional. In such a representation, the matrices $\tilde{\gamma}_1$ and $\tilde{\gamma}_2$ already span the full matrix algebra, leaving no room for a non-zero nilpotent matrix that anticommutes with both. Consequently, no faithful inhomogeneous two-dimensional representation of the full Carroll Clifford algebra exists in three dimensions.

The minimal resolution is to double the representation space. A convenient construction is obtained by taking the representation of 4d Carroll-Dirac matrices and dropping one of the spatial gamma matrices. The following choice of Dirac matrices and conjugation matrix $\Lambda$ will be convenient later on:
\begin{align}
     \tilde{\Gamma}_0= \begin{pmatrix} 0 & 0\\ \gamma_0 & 0 \end{pmatrix}\,,\quad
    \tilde{\Gamma}_i= \begin{pmatrix} \gamma_i & 0\\ 0 & \gamma_i \end{pmatrix}\,,\quad
    \Lambda = \begin{pmatrix} 0 & -\ii \gamma_0 \\ -\ii \gamma_0 & 0  \end{pmatrix}
    \label{3dcliff}
\end{align}
where
\begin{eqn}
    \gamma_{0}=-\ii \sigma^2, \quad \gamma_1=\sigma^1,\quad \gamma_2=\sigma^3.
    \label{relativistic3dgamma}
\end{eqn}
are relativistic 3d Dirac matrices satisfying $\{\g^{\mu},\g^{\nu}\} = 2\eta^{\mu\nu}$, with  $\eta^{\mu\nu} = \text{diag}(-1,1,1)$. One can easily verify that these matrices satisfy the Carroll Clifford algebra. 
Hence, the minimal faithful representation in $\mathcal{R}^{\downarrow}_I$ of the Carroll Clifford algebra in three spacetime dimensions is four-dimensional.

Next let us see how to obtain the action for 3d Carrollian fermions from the $c \rightarrow 0$ limit of relativistic Dirac fermions. This will serve as a toy model for the Carrollian limit of the ABJM theory considered in the next section. We start with the relativistic action
\begin{equation}
S_{\text{3d}}^{\text{Rel}} =
\int d^3x \,
\ii \bar{\psi} \gamma^\mu \partial_\mu \psi ,
\end{equation}
where the $\g$'s are the relativistic gamma matrices in \eqref{relativistic3dgamma} and $\bar{\psi}=\psi^{\dagger} \gamma^0$.   Note that the spinors have two compnents but we will suppress spinor indices in most of what follows. We then take the Carroll limit by rescaling $t\rightarrow c t$ and taking $c\rightarrow 0$, which suppresses the spatial derivatives and yields
\begin{equation}
S_{\text{3d}}^{\text{Carroll}} =
\int d^3x \, \ii \bar{\psi} \gamma^0 \partial_t \psi .
\end{equation}
We can then make the action manifestly Carroll invariant using the Dirac matrices in \eqref{3dcliff}:
\begin{equation}
\label{eq:ferm_gauge_kin_rescale}
S_{3d}^{\text{Carroll}} =
\int d^3x \,
\bar\Psi \tilde{\Gamma}_0 \partial_t \Psi ,
\end{equation}
where 
\begin{equation}
\Psi=
\begin{pmatrix}
\psi \\
0
\end{pmatrix}
\label{3dcspinor}
\end{equation}
and $\bar \Psi = \Psi^\dagger \Lambda$. From the explicit block structure of $\tilde{\Gamma}_0$ in \eqref{3dcliff}, one sees that the action depends only on the upper component $\psi$ of $\Psi = (\psi\,,\chi)^T$ so we can set the lower component to zero. Under Carroll boost the spinor and its conjugate transform as 
\begin{align}\label{3dcarrollboost}
    &\delta\Psi = b^ix_i\partial_t\Psi + \tilde{\Sigma}_{0i}\Psi \nonumber\\
    &\delta\bar{\Psi} = b^ix_i\partial_t\bar{\Psi} - \bar{\Psi}\tilde{\Sigma}_{0i}\,,
\end{align}
where $\tilde{\Sigma}_{0i} = \frac{1}{4}[\tilde{\Gamma}_0, \tilde{\Gamma}_i]$ is the spin part of the Carroll boost generator. The $b^ix_i\partial_t$ term contributes a total time derivative to the Lagrangian. The spin term contributes a term proportional to $[\tilde{\Sigma}_{0i},\tilde{\Gamma}_0]= 0$, which vanishes by the Clifford algebra, confirming invariance of the Lagrangian up to a total derivative: 
\begin{align}
    \delta\mathcal{L}_{3d}^{\text{Carroll}} = \partial_t\left(b^ix_i\mathcal{L}_{3d}^{\text{Carroll}}\right)\,.
\end{align}
Hence the action in \eqref{eq:ferm_gauge_kin_rescale} is invariant under Carroll boosts. 

\subsection{A first look at Carroll SUSY: A toy model}
We now construct the simplest example of a supersymmetric Carrollian field theory. As we mentioned earlier, a key feature of Carrollian systems is that fermionic degrees of freedom are reduced relative to their relativistic counterparts. As a result, matching bosonic and fermionic degrees of freedom in a supersymmetric theory requires twice as many Carroll fermions as bosons. 

The minimal supersymmetric multiplet therefore consists of one real Carroll scalar and two real Carroll fermions. We consider a single real scalar field $\phi(x,t)$ together with two real Carroll fermions $\psi_L(x,t)$ and $\chi_R(x,t)$, transforming in the representation $\mathcal{R}^\downarrow_I$. The action is given by, 
\begin{align}
    \mathcal{S} 
    =\int d^dx\,dt\,\left[ \frac{1}{2}(\partial_t \phi)^2 
    + \ii\bigl(\psi_L \partial_t \psi_L + \chi_R \partial_t \chi_R\bigr)\right],
\end{align}
which contains only time derivatives, as expected in a (electric) Carrollian theory. It is convenient to combine the two real fermions into a single complex fermion,
\begin{align}
    \Psi = \frac{1}{\sqrt{2}}(\psi_L + \ii\chi_R), 
    \qquad 
    \Psi^\dagger = \frac{1}{\sqrt{2}}(\psi_L - \ii\chi_R).
\end{align}
In terms of these variables, the Lagrangian becomes
\begin{align}
    \mathcal{L} 
    = \frac{1}{2}(\partial_t \phi)^2 
    + \ii\,\Psi^\dagger \partial_t \Psi .
\end{align}
The canonical momenta conjugate to the fields are
\begin{align}
    \pi_\phi = \frac{\partial \mathcal{L}}{\partial \dot{\phi}} = \dot{\phi},\quad \pi_\Psi = \ii\,\Psi^\dagger.
\end{align}
The fermionic sector is first order in time derivatives and gives rise to second-class constraints. After eliminating these constraints and passing to Dirac brackets, the non-vanishing equal-time brackets are
\begin{align}
    \{\phi(t,\vec{x}), \pi_\phi(t,\vec{x}')\} = i\delta^{(d)}(\vec{x}-\vec{x}'),
    \qquad
    \{\Psi(t,\vec{x}), \Psi^\dagger(t,\vec{x}')\} = \delta^{(d)}(\vec{x}-\vec{x}').
\end{align}
The Hamiltonian density is simply
\begin{align}
    \mathcal{H} = \frac{1}{2}\pi_\phi^2.
\end{align}
Under supertranslations $ t \to t + f(x)$, the fields transform as
\begin{subequations}
    \begin{align}
    \delta_f \phi &= \{\phi, M_f\} = f(x)\,\dot{\phi}, \\
    \delta_f \Psi &= \{\Psi, M_f\} = f(x)\,\dot{\Psi}, \\
    \delta_f \Psi^\dagger &= \{\Psi^\dagger, M_f\} = f(x)\,\dot{\Psi}^\dagger .
\end{align}
\end{subequations}
where $M_f = \int d^d x\, f(x)\,\mathcal{H}$. 
Since the Lagrangian contains only time derivatives, these variations leave the action invariant up to a total time derivative, confirming that the model is invariant under the full infinite-dimensional supertranslation symmetry.

\paragraph{Supercharges and transformations.}
The theory admits a pair of conserved fermionic charges,
\begin{align}
    Q = \int d^dx\,\pi_\phi(t,\vec{x})\, \Psi^\dagger(t,\vec{x}), 
    \qquad 
    Q^\dagger = \int d^dx\,\pi_\phi(t,\vec{x})\, \Psi(t,\vec{x}),
\end{align}
which generate supersymmetry transformations via
\begin{align}
    \delta_\epsilon(\,\cdot\,) 
    = \bigl\{\,\cdot,\; \epsilon Q + \epsilon^\dagger Q^\dagger \bigr\},
\end{align}
where $\epsilon$ and $\epsilon^\dagger$ are Grassmann-valued parameters. Acting on the fundamental fields, one finds
\begin{align}
    \delta \phi &= \epsilon\, \Psi^\dagger + \epsilon^\dagger\, \Psi, \quad
    \delta \Psi = -\epsilon^\dagger\, \dot{\phi}, 
    \quad 
    \delta \Psi^\dagger = -\epsilon\, \dot{\phi}.
\end{align}
Two successive supersymmetry transformations close onto a time translation. Using the canonical brackets, the supercharges satisfy
\begin{align}
    \{Q, Q^\dagger\} = 2H,
    \qquad
    \{Q, Q\} = \{Q^\dagger, Q^\dagger\} = 0,
    \qquad
    [H, Q] = [H, Q^\dagger] = 0.
\end{align}
This realizes a Carrollian supersymmetry algebra in which the Hamiltonian appears as the central bosonic generator.

Finally, expressing the transformations in terms of the original real fermions and real Grassmann parameters
\begin{align}
    \eta_1 = \frac{1}{\sqrt{2}}(\epsilon + \epsilon^\dagger),
    \qquad 
    \eta_2 = \frac{\ii}{\sqrt{2}}(-\epsilon + \epsilon^\dagger),
\end{align}
the supersymmetry variations take the form
\begin{align}
    \delta \phi &= \eta_1 \psi_L - \eta_2 \chi_R, \quad 
    \delta \psi_L = -\eta_1\, \dot{\phi},
    \quad \delta \chi_R = -\eta_2\, \dot{\phi}.
\end{align}
This explicitly exhibits the doubling of Carroll fermions required to form a supersymmetric multiplet with a single bosonic degree of freedom. Moreover, the supertranslation generators commute with the supercharges,
\begin{align}
    [M_f, Q] = [M_f, Q^\dagger] = 0,
\end{align}
so that the supersymmetry algebra extends consistently to a
super-Carrollian algebra with supertranslations.

At first sight, the model constructed above resembles standard $\mathcal{N}=1$ supersymmetric quantum mechanics. 
However, the two theories are conceptually distinct. The present model is defined intrinsically on a $d+1$-dimensional flat Carrollian spacetime and realizes a contraction of the super-Poincaré algebra in the ultra-relativistic limit. While the supersymmetry algebra takes the same schematic form as in supersymmetric quantum mechanics, its representation theory and geometric interpretation are fundamentally Carrollian.

\section{ABJM Theory and Its Carroll Limit} \label{abjmcar}

As we stated in the introduction, a promising strategy to construct a concrete example of flat space holography, is to take the flat space limit of a known example of AdS/CFT. This would involve taking the flat space limit of quantum gravity in the bulk and the Carrollian limit of the conformal field theory in the boundary. Initial steps in this direction have been taken for the AdS$_4$/CFT$_3$ correspondence, where the bulk theory is M-theory in AdS$_4 \times$S$^7/\mathbb{Z}_{k}$ and the boundary theory is a superconformal Chern-Simons theory with bi-adjoint matter known as the ABJM theory \cite{Aharony:2008ug}. In \cite{Lipstein:2025jfj} the Carrollian limit of 4-point correlators of protected operators in the ABJM theory were shown to describe supergravity amplitudes with four-dimensional kinematics, building on previous results in \cite{Alday:2024yyj,Alday:2020dtb}. At the same time, the Carrollian limit of Chern-Simons theories coupled to scalar matter were analysed in \cite{Bagchi:2024efs} and shown to exhibit extended BMS$_4$ symmetry, as expected for a putative dual to 4d asymptotically flat space.

What was missing in that analysis was a treatment of fermions. Since the fermions start off as 3d relativistic Dirac spinors, it was not obvious at the time that they could be interpreted as Carrollian fermions after taking the speed of light to zero. The purpose of this section is to show that this is indeed the case. This is especially nontrivial given that the 3d Carrollian Clifford algebra is actually four-dimensional, as explained in the previous section. In Appendix \ref{app:comp} we show that the Carrollian ABJM action enjoys extended BMS$_4$ symmetry (generalising the analysis in \cite{Bagchi:2024efs} to include fermions) and in the next section, we show that the it enjoys a supersymmetric extension of this infinite-dimensional symmetry.

\subsection{Relativistic ABJM}

The ABJM theory is a 3d Chern-Simons Matter (CSM) theory with classical $\mathcal{N}=6 $ supersymmetry and gauge group ${U(N)_{k}} \times {U(N)_{-k}}$, where the subscripts denote Chern-Simons levels. The matter consists of complex scalars $X_A$ and Dirac fermions $\psi_A$ transforming in the bifundamental representation of the gauge group, where $A=1\ldots 4$ are fundamental indices of the $SU(4)$ R-symmetry and we do not display spinor or color indices. The adjoint fields will be denoted by upper R-symmetry indices, i.e. $X^A$ and $\psi^A$. The explicit Lagrangian for this theory was first constructed in \cite{Benna:2008zy} and \cite{abjm_sym} using a superfield and component formalism, respectively. In this paper we will use the component formalism. 

Following \cite{abjm_sym}, the ABJM action can be written as 
\begin{eqn}\label{eq:abjm_net}
S&= S_{\text{kin}}+S_{\text{CS}}+S_{\psi^2 X^2}+S_{X^6}\,\\
&= \frac{k}{2\pi}\int d^3x\,\left(\mathcal{L}_{\text{kin}}+\mathcal{L}_{\text{CS}}+\mathcal{L}_{\psi^2 X^2}+\mathcal{L}_{X^6}\right).
\end{eqn}
The kinetic terms for the matter fields are given by
\begin{eqn}\label{eq:abjm_kin}
\mathcal{L}_{\text{kin}}= \text{Tr}\Bigl( -D_{\mu}X^{ A}D^{\mu}X_A + \ii \bar{\psi}_{ A}\gamma^{\mu}D_{\mu}\psi^A \Bigr) 
\end{eqn}
where $\gamma^{\mu}$ are the relativistic 3d Dirac matrices given in \eqref{relativistic3dgamma} and $\bar{\psi}^A= (\psi^A)^{\rm{T}} \gamma^0$. The covariant derivative is given by 
\begin{eqn}
D_{\mu}X_A &=\partial_{\mu}X_A +\ii \Bigl(A_{\mu} X_A-X_A \hat{A}_{\mu} \Bigr)\\
D_{\mu}\psi_A &=\partial_{\mu}\psi_A +\ii\Bigl(\hat{A}_{\mu}\psi_A - \psi_A A_{\mu}\Bigr) .\\
\end{eqn}
The Chern-Simons action reads
\begin{eqn}\label{eq:abjm_cs}
\mathcal{L}_{\text{CS}}&= \frac{1}{2} \varepsilon^{\mu\nu\rho}\text{Tr}\Bigl(A_{\mu}\partial_{\nu}A_{\rho}+\frac{2\ii}{3}A_{\mu}A_{\nu}A_{\rho}-\hat{A}_{\mu}\partial_{\nu}\hat{A}_{\rho}-\frac{2\ii}{3}\hat{A}_{\mu}\hat{A}_{\nu}\hat{A}_{\rho}  \Bigr),
\end{eqn}
where $A_{\mu}$ and $\hat{A}_{\mu}$ are gauge fields of the two $U(N)$ gauge groups with opposite Chern-Simons levels $\pm k$. Moreover the Yukawa interactions are 
\begin{eqn}\label{eq:abjm_Yukawa}
\mathcal{L}_{\psi^2 X^2} &= \ii\, \text{Tr}\Bigl( \varepsilon^{ABCD}\Bigl(\bar{\psi}_A X_B \psi_C X_D\Bigr)-\varepsilon_{ABCD}\Bigl(\bar{\psi}^{ A}X^{ B}\psi^{ C}X^{ D} \Bigr) \\&+ \bar{\psi}^{ A}\psi_A X_B X^{ B}-\bar{\psi}_A\psi^{ A}X^{ B}X_{B} +2\bar{\psi}_A \psi^{ B} X^{ A}X_B- 2\bar{\psi}^{ B}\psi_A X_BX^{ A}     \Bigr)
\end{eqn} 
with $\varepsilon^{1234}=\varepsilon_{1234}=1$, and the scalar potential takes the form
\begin{eqn}\label{eq:abjm_pot}
\mathcal{L}_{X^6} &=\frac{1}{3}\text{Tr}\Bigl(X^{ A}X_A X^{ B}X_{B}X^{ C}X_C + X_AX^{ A} X_{B}X^{ B}X_CX^{ C}\\& +4  X_A X^{ B}X_C X^{ A}X_{B}X^{ C}  - 6X^{ A} X_{ B}X^{B}X_A X^{ C}X_C \Bigr) \,.
\end{eqn}
The action is invariant under the following supersymmetry transformations:
\begin{eqn}\label{eq:abjm_susy_trans}
\delta X_A &= \ii  C^I_{AB}\bar{\psi}^{ B}\epsilon^I\,,\quad 
\delta X^{ A} = -\ii \tilde{C}^{IAB}\bar{\psi}_B\epsilon^I\\
\delta \psi_A &= C^I_{AB} \gamma^{\mu}\epsilon^I D_{\mu}X^{ B} +N^I_A \epsilon^I \,,\quad
\delta \psi^{ A} =-\tilde{C}^{IAB}\gamma^{\mu}\epsilon^I D_{\mu}X_B +N^{IA}\epsilon^I \\
\delta A_{\mu} &= -C^I_{AB}\bar{\psi}^{ A}\gamma_{\mu}\epsilon^I X^{ B}- \tilde{C}^{IAB}X_B \bar{\psi}_A \gamma_{\mu}\epsilon^I\,, \\ 
\delta \hat{A}_{\mu} &=  -C^I_{AB}X^{ B}\bar{\psi}^A\gamma_{\mu}\epsilon^{ I} - \tilde{C}^{IAB} \bar{\psi}_A  \gamma_{\mu}\epsilon^I X_B\\
\end{eqn}
where $I=1\ldots 6$ labels the $\mathfrak{so}(6)$ R-symmetry, $\epsilon^I$ are Majorana fermionic supersymmetry variation parameters, and $C^I_{AB},\tilde{C}^{IAB} $ are Clebsch-Gordan coefficients which convert the $\mathfrak{su}(4) $ representation $\mathbf{6} $ into $\mathbf{4}\otimes \mathbf{4} $ such that $C^I_{BA}=-C^I_{AB} $, $\tilde{C}^I_{AB}=-\Bigl(C^I_{AB}\Bigr)^{*} $, $\frac{1}{2}\varepsilon^{ABCD}C^I_{CD}=\tilde{C}^{IAB} $ and $C^I_{AB}\tilde{C}^{JBC} + C^J_{AB}\tilde{C}^{IBC}=2\delta^{IJ}\delta_A{}^B $. The terms $ N^{IA}$ and $N^I_A $ are given by
\begin{eqn}
N^{IA} &=\tilde{C}^{IAB}\Bigl(X_C X^{ C} X_B -X_B  X^{ C} X_C \Bigr) -2\tilde{C}^{IBC} X_B X^{ A}X_C\\
N^I_A &= C^I_{AB}\Bigl(X^{ C}X_CX^{ B} - X^{ B} X_C X^{ C}\Bigr) -2C^I_{BC}X^{ B}X_A X^{ C}\,.
\end{eqn}

The ABJM theory has three-dimensional conformal symmetry which combines with the $\mathcal{N}=6$ supersymmetry to give superconformal group $OSp(6|4)$. The superconformal transformations are obtained by replacing $\epsilon^I $ with $\gamma^{\nu}x_{\nu}\xi^I$ above, and adding compensating terms in the fermion variations:
\begin{eqn}\label{eq:abjm_susy_conf_trans}
\delta X_A &= \ii C^I_{AB}\bar{\psi}^{ B}\gamma^{\nu}x_{\nu} \xi^I\,,\quad
\delta X^{ A} = -\ii \tilde{C}^{IAB}\bar{\psi}_B\gamma^{\nu}x_{\nu} \xi^I \\
\delta \psi_A &= C^I_{AB} \gamma^{\mu}\gamma^{\nu}x_{\nu}\xi^I D_{\mu}X^{ B} +N^I_A \gamma^{\nu}x_{\nu}\xi^I + C^I_{AB}X^{ B}\xi^I \\
\delta \psi^{ A}&=-\tilde{C}^{IAB}\gamma^{\mu}\gamma^{\nu}x_{\nu}\xi^I D_{\mu}X_B +N^{IA}\gamma^{\nu}x_{\nu}\xi^I -\tilde{C}^{IAB}X_B \xi^I \\
\delta A_{\mu} &= -C^I_{AB}\bar{\psi}^{ A}\gamma_{\mu}\gamma^{\nu}x_{\nu}\xi^I X^{ B}- \tilde{C}^{IAB}X_B \bar{\psi}_A \gamma_{\mu}\gamma^{\nu}x_{\nu}\xi^I\\
\delta \hat{A}_{\mu} &=  -C^I_{AB}X^{ B}\bar{\psi}^{ A}\gamma_{\mu}\gamma^{\nu}x_{\nu}\xi^I - \tilde{C}^{IAB} \bar{\psi}_A  \gamma_{\mu}\gamma^{\nu}x_{\nu}\xi^I X_B\,.\\
\end{eqn}
Finally, the theory is also invariant under the R-symmetry variations
\begin{eqn}\label{eq:abjm_R_sym}
\delta X_A &=\frac{1}{2} \alpha^{IJ} C^{IJ}{}_A{}^B X_B\,,\delta X^A = -\frac{1}{2} \alpha^{IJ} C^{IJ}{}_B{}^A X^B \\
\delta \psi_A &=\frac{1}{2} \alpha^{IJ} C^{IJ}{}_A{}^B \psi_B\,,\delta \psi^A = -\frac{1}{2} \alpha^{IJ} C^{IJ}{}_B{}^A \psi^B\\
\delta A_{\mu} &= \delta \hat{A}_{\mu} =0 \\
\end{eqn}
where $C^{IJ}{}_A{}^B=\frac{1}{2}(C^I_{AC}\tilde{C}^{JCB}-C^J_{AC}\tilde{C}^{ICB}) $ and $\alpha^{IJ}$ are infinitesimal parameters satisfying $\alpha^{IJ}=-\alpha^{JI}$.

\subsection{Carroll Limit}

Following the strategy adopted in \cite{carroll_csm}, the electric Carrollian limit of the ABJM theory can be obtained by rescaling the spacetime coordinates and fields with appropriate factors of $c$ and keeping only terms of order $\mathcal{O}(c^0)$. Requiring that the leading contributions to the action remain finite and contain kinetic terms for all the matter fields in the limit $c\to 0$ limit uniquely fixes the scaling of the fields: 
\begin{eqn}\label{eq:abjm_c_rescale}
x^0 &\rightarrow ct \,,\quad
d^3x \rightarrow cd^3x\\
A_{0}&\rightarrow c^{-1}A_{t}\,,\quad A_{i}\rightarrow A_{i}\\
\hat{A}_{0}&\rightarrow c^{-1}\hat{A}_{t}\,,\quad\hat{A}_{i}\rightarrow\hat{A}_{i}\\
X_A &\rightarrow c^{1/2} X_A\,,\quad
X^{ A} \rightarrow c^{1/2} X^{ A}\\
\psi_A &\rightarrow \psi_A \,,\quad
\psi^{ A} \rightarrow \psi^{ A}\,\\
\epsilon^I &\rightarrow c^{1/2}\epsilon^I\,,\quad
\xi^I \rightarrow c^{1/2}\xi^I\,, \quad \alpha^{IJ}\rightarrow \alpha^{IJ}.
\end{eqn}

Under the rescaling, the Chern-Simons action remains unchanged and the kinetic terms for the matter fields become
\begin{eqn}\label{eq:abjm_kin_rescaled}
S_{\text{kin}}&\rightarrow \frac{k}{2\pi} \int d^3x\text{Tr}\Bigl(D_tX^{ A}D_t X_A  + \ii\, \bar{\psi}_{ A}\gamma^0 D_t \psi^A \Bigr).  \\
\end{eqn}
Note that the spatial derivatives are suppressed in this limit while $S_{\psi^2X^2}$ and $S_{X^6} $ vanish since they are $\mathcal{O}(c^2)$ and $\mathcal{O}(c^4)$, respectively. The action \eqref{eq:abjm_net} therefore reduces to
\begin{eqn}\label{eq:abjm_net_rescaled}
S_{\text{Carroll}} &= S_{\text{kin}} + S_{\text{CS}} \\
&=\frac{k}{2\pi} \int d^3x \text{Tr}\Bigl( D_tX^{ A}D_t X_A  +  \ii\, \bar{\psi}_{ A}\gamma^0 D_t \psi^A   \\& \quad + \frac{1}{2}\varepsilon^{\mu\nu\rho}(A_{\mu}\partial_{\nu}A_{\rho}+\frac{2\ii}{3}A_{\mu}A_{\nu}A_{\rho}-\hat{A}_{\mu}\partial_{\nu}\hat{A}_{\rho}-\frac{2\ii}{3}\hat{A}_{\mu}\hat{A}_{\nu}\hat{A}_{\rho}) \Bigr) \\
\end{eqn}
which leads to the following equations of motion:
\begin{eqn}\label{eq:eom_abjm_rescaled}
&D_t^2 X_A =0 \, , \gamma^0 D_t \psi_A =0, \\
&F_{ti} =0 \, , \hat{F}_{ti} =0, \\
&F_{12} =\ii (D_t X_A X^A - X_A D_t X^A)- \bar{\psi}^A\gamma^0 \psi_A,  \\
&\hat{F}_{12}=\ii (X^A D_t X_A -D_t X^A X_A) + \bar{\psi}_A \gamma^0 \psi^A. \\
\end{eqn}
After rescaling by $c$ and taking $c\rightarrow 0$ as described above, the supersymmetry variations in \eqref{eq:abjm_susy_trans} reduce to
\begin{eqn}\label{eq:abjm_susy_trans_rescaled}
\delta X_A &=\ii C^I_{AB}\bar{\psi}^{ B}\epsilon^I\,,\quad
\delta X^{ A} = -\ii \tilde{C}^{IAB}\bar{\psi}_B\epsilon^I\\
\delta \psi_A &= C^I_{AB} \gamma^{0}\epsilon^I D_{t}X^{ B} \,,\quad
\delta \psi^{ A} =-\tilde{C}^{IAB}\gamma^{0}\epsilon^I D_{t}X_B \\
\delta A_{i} &= \delta \hat{A}_{i}= \delta A_{t}= \delta \hat{A}_t= 0\,.
\end{eqn}
Similarly the superconformal variations in \eqref{eq:abjm_susy_conf_trans} become
\begin{eqn}\label{eq:abjm_susy_conf_trans_rescaled}
\delta X_A &= \ii C^I_{AB}\bar{\psi}^{ B}\gamma^{i}x_{i} \xi^I\,,\quad
\delta X^{ A} = -\ii \tilde{C}^{IAB}  \bar{\psi}_B\gamma^{i}x_{i} \xi^I\\
\delta \psi_A &= C^I_{AB} \gamma^{0}(\gamma^{i}x_{i} \xi^I )D_{t}X^{ B} \,,\quad
\delta \psi^{ A} =-\tilde{C}^{IAB}\gamma^{0}\gamma^{i}x_{i} \xi^I D_{t}X_B   \\
\delta A_{i} &= \delta \hat{A}_{i}= \delta A_{t}= \delta \hat{A}_t= 0\,.
\end{eqn}
Finally, the R-symmetry variation in \eqref{eq:abjm_R_sym} remain the same under rescaling:
\begin{eqn}\label{eq:abjm_R_sym_rescaled}
\delta X_A &=\frac{1}{2} \alpha^{IJ} C^{IJ}{}_A{}^B X_B\,,\delta X^A = -\frac{1}{2} \alpha^{IJ} C^{IJ}{}_B{}^A X^B \\
\delta \psi_A &=\frac{1}{2} \alpha^{IJ} C^{IJ}{}_A{}^B \psi_B\,,\delta \psi^A = -\frac{1}{2} \alpha^{IJ} C^{IJ}{}_B{}^A \psi^B\\
\delta A_{\mu} &= \delta \hat{A}_{\mu} =0 \\
\end{eqn}

A notable feature is that after taking the Carrollian limit the gauge fields do not transform under either supersymmetry or superconformal variations. At first glance, it might seem to be in tension with the expectation that the commutator of two supersymmetry variations should produce a (gauge-covariant) translation. This apparent tension can be resolved by noting that such a variation of the gauge fields vanishes after applying the equations of motion of the Carrollian theory. In more detail, if we take the commutator of two supersymmetry variations we would expect to obtain (based on \eqref{qq})
\begin{eqn}
[\delta_1,\delta_2] A_t &= 2\ii\, \bar{\epsilon}^I_1\gamma^0\epsilon^I_2 F_{tt} =0 \\
[\delta_1,\delta_2] A_i &= 2\ii\, {\epsilon}^I_1\gamma^0{\epsilon}^I_2 F_{ti}=0,\\
\end{eqn}
with similar expressions for hatted gauge fields. Note that the first line vanishes by the antisymmetry of the field strength while the the second line vanishes due to the equations of motion \eqref{eq:eom_abjm_rescaled}. The commutator of two supersymmetries can also yield a gauge transformation \cite{Bagger:2007jr}, but this contribution is subleading in the $c\rightarrow0$ limit after performing the recalings in \eqref{eq:abjm_c_rescale}. Similarly,
based on \eqref{qs} and \eqref{ss} we expect to obtain
\begin{eqn}
[\delta_1, \tilde{\delta}_2] A_t &= 2\ii \bar{\epsilon}^I_1 \gamma^0 \gamma^j \xi_2^I x_j F_{tt}=0 \\
[\delta_1, \tilde{\delta}_2] A_i &= 2\ii \bar{\epsilon}^I_1 \gamma^0 \gamma^j \xi_2^I x_j F_{ti}=0\\
[\tilde{\delta}_1, \tilde{\delta}_2] A_t &= 2\ii \bar{\xi}^I_1 \gamma^0  \xi_2^I \vec{x}^2 F_{tt}=0\\
[\tilde{\delta}_1, \tilde{\delta}_2] A_i &= 2\ii \bar{\xi}^I_1 \gamma^0  \xi_2^I \vec{x}^2 F_{ti}=0,\\
\end{eqn}
where $\tilde{\delta}$ denotes the superconformal variation in \eqref{eq:abjm_susy_conf_trans_rescaled} and $\delta$ denotes the supersymmetry variation in \eqref{eq:abjm_susy_trans_rescaled}. Hence, commutators of Carrollian supersymmetric/superconformal variations of the gauge fields vanish in a way consistent with the Carrollian superconformal algebra.

\subsection{Recast in terms of Carroll Dirac Matrices}

At the moment it is not obvious that the $c\rightarrow 0$ limit of the ABJM theory in \eqref{eq:abjm_net_rescaled} enjoys Carrollian conformal symmetry because it is written in terms of relativistic 3d Dirac matrices. In order to make this symmetry manifest, let us write the action in terms of Carrollian Dirac matrices in \eqref{3dcliff}:
\begin{eqn}\label{eq:abjm_carroll}
S_{\text{Carroll}} &= S_{\text{CS}}+ \frac{k}{2\pi} \int d^3x \text{Tr}\Bigl( D_t X^{A}D_t X_A + \bar{\Psi}_A \tilde{\Gamma}_{0} D_t \Psi^A \Bigr), \\
\end{eqn}
where $S_{\text{CS}}$ is given in \eqref{eq:abjm_cs},
\begin{eqn}
\Psi^A &= \begin{pmatrix}  \psi^A \\ 0 \end{pmatrix}\,,\quad
\bar{\Psi}_A = (\Psi_A)^{\rm{T}} \Lambda \,,\quad
\Psi_A = \begin{pmatrix}  \psi_A \\ 0 \end{pmatrix}\,,
\end{eqn}
and the covariant derivative is defined as 
\begin{align}
    D_t\Psi_A = \partial_t \Psi_A + \ii (\hat{A}_t \Psi_A - \Psi_A A_t  )\,.
\end{align}
Following the discussion below \eqref{eq:ferm_gauge_kin_rescale}, it is strightforward to verify that the action above is invariant under Carrollian boosts. 

Moreover, it is straghtforward to recast the Carrollian supersymmetry transformations in \eqref{eq:abjm_susy_trans_rescaled} in terms of Carrollian Dirac matrices by defining the Carrollian supersymmetry parameter as 
\begin{align}
    \mathcal{E}^I=\begin{pmatrix} 0 \\ \epsilon^{I} \end{pmatrix}. 
\end{align}
The supersymmetry transformations then read
\begin{eqn}\label{eq:abjm_carroll_susy_trans}
\delta X_A &= -C^I_{AB} \bar{\Psi}^B \mathcal{E}^I\,,\quad
\delta X^{A} = \tilde{C}^{IAB}  \bar{\Psi}_B\mathcal{E}^I \\
\delta \Psi_A &=  C^I_{AB} \tilde{\Gamma}^{\dagger}_{0}\mathcal{E}^I D_t X^{B} \,,\quad
\delta \Psi^A = -\tilde{C}^{IAB}  \tilde{\Gamma}^{\dagger}_{0}\mathcal{E}^I D_t X_B.
\end{eqn}
Note that $\mathcal{E}^I$ satisfies the following Carrollian Majorana condition:
\begin{align}
    \left(\mathcal{E}^{I}\right)^{\dagger}\Lambda = (\mathcal{E}^I)^T\mathcal{C},
\label{3dmajorana}
\end{align}
where $\mathcal{C}$ is the charge conjugation matrix 
\begin{align}
    \mathcal{C} = \begin{pmatrix}
        0 & -\sigma^2\\
        -\sigma^2 & 0
    \end{pmatrix}.
\end{align}
The charge conjugation matrix satisfies the following properties 
\begin{align}
    \mathcal{C}^T = - \mathcal{C}\,,\quad  \mathcal{C}^\dagger =  \mathcal{C}^{-1}\,,\quad  \mathcal{C}\tilde{\Gamma}_{\mu} \mathcal{C}^{-1} = - (\tilde{\Gamma}_{\mu})^T\,.
\end{align} 
See \cite{Bagchi:2026lgk} for further discussion of charge conjugation and the Majorana condition in the Carrollian context.  

In a similar manner, one can write the superconformal variations in \eqref{eq:abjm_susy_conf_trans_rescaled} in a manifestly covariant way. Defining the superconformal parameter 
\begin{equation}
\Xi^I = \begin{pmatrix} 0 \\ \xi^{I} \end{pmatrix} 
\end{equation}
the global superconformal variations are given by
\begin{eqn}\label{eq:abjm_carroll_susy_conf_trans}
\delta X_A &= -C^I_{AB} \bar{\Psi}^B \Gamma_i x^i \Xi^I\,,\quad
\delta X^{A} = \tilde{C}^{IAB} \bar{\Psi}_B \Gamma_i x^i \Xi^I \\
\delta \Psi_A &=  C^I_{AB} \tilde{\Gamma}^{\dagger}_{0} \Gamma_i x^i \Xi^I D_t X^{*B} \,,\quad
\delta {\Psi}^A = -\tilde{C}^{IAB} \tilde{\Gamma}^{\dagger}_{0} \Gamma_i x^i  \Xi^I D_t X_B \,.
\end{eqn}
It was shown in \cite{carroll_csm} that the bosonic sector of the action in \eqref{eq:abjm_carroll} is invariant under the extended BMS$_4$ symmetry reviewed in section \eqref{bosonicsymm} and \ref{carrolcftrep}. To establish the extended BMS invariance of the full Carrollian action \eqref{eq:abjm_carroll}, it is therefore sufficient to examine only the fermionic sector. We work this out in Appendix \ref{app:comp}.

\subsection{Propagators} \label{props}
To give a flavour of the quantum mechanical aspects of the theory, we now compute propagators. For simplicity, we will work in temporal gauge $A_t=\hat{A}_t=0$. The Carrollian ABJM action in \eqref{eq:abjm_net_rescaled} then reduces to
\begin{eqn}\label{eq:abjm_net_rescaled_axial_gauge}
S_{\text{Carroll}}  &=\frac{k}{2\pi} \int d^3x \text{Tr}\Bigl( \partial_tX^{ A}\partial_t X_A  +  \ii\, \bar{\psi}_{ A}\gamma^0 \partial_t \psi^A    - \frac{1}{2}\varepsilon^{ij}(A_{i}\partial_{t}A_{j}-\hat{A}_{i}\partial_{t}\hat{A}_{j}) \Bigr) \\
\end{eqn}
The 2-point functions can be easily deduced from the Dyson-Schwinger equations of motion: 
\begin{subequations}
\begin{align}
-\frac{k}{2\pi} \partial_t^2\langle X_A{}_m{}^{\hat{m}}(t,\vec{x}) X^B{}_{\hat{n}}{}^n(0,\vec{0}) \rangle &= \ii \delta_A^B \delta^{\hat{m}}_{\hat{n}} \delta_{m}^{n} \delta(t)\delta^2(\vec{x})\\
\frac{k}{2\pi} \gamma^0\partial_t\langle {(\psi^A)}_m{}^{\hat{m}}(t,\vec{x}) {(\bar{\psi}_{B})}_{\hat{n}}{}^n(0,\vec{0}) \rangle&= \delta^A_B\delta^m_n\delta_{\hat{m}}^{\hat{n}}\delta(t)\delta^2(\vec{x}) \\
\ii\frac{k}{2\pi} \varepsilon^{ik} \partial_t \langle A_k{}_{m}{}^n(t,\vec{x}) A_j{}_{p}{}^{q}(0,\vec{0}) \rangle &=\delta^i_j \delta_m^q \delta_n^p \delta(t)\delta^2(\vec{x}) \\
- \ii\frac{k}{2\pi} \varepsilon^{ik} \partial_t \langle \hat{A}_k{}_{\hat{m}}{}^{\hat{n}}(t,\vec{x}) \hat{A}_j{}_{\hat{p}}{}^{\hat{q}}(0,\vec{0}) \rangle & = \delta^i_j \delta_{\hat{m}}^{\hat{q}} \delta_{\hat{n}}^{\hat{p}}   \delta(t)\delta^2(\vec{x}), 
    \end{align}
\end{subequations}
which are solved by
\begin{subequations}
    \begin{align}
\langle X_{A}{}_{m}{}^{\hat{m}}(t,\vec{x})X^{B}{}_{\hat{n}}{}^{n}(0,\vec{0})\rangle&=-\ii\delta_{A}^{B}\delta_{\hat{n}}^{\hat{m}}\delta_{m}^{n}\frac{\pi}{k}|t|\delta^{2}(\vec{x})\\
\langle{\psi^A}{}_{\hat{m}}{}^{m}(t,\vec{x}){\bar{\psi}_B}{}_{n}{}^{\hat{n}}(0,\vec{0})\rangle &=-\delta^A_B\delta^m_n\delta_{\hat{m}}^{\hat{n}} \gamma^0 \frac{\pi}{k}\text{sgn}(t)\delta^{2}(\vec{x})\\
\langle A_{i}^{M}(t,\vec{x})A_{j}^{N}(0,\vec{0})\rangle &=\ii\delta^{MN}\varepsilon_{ij}\frac{\pi}{k}\text{sgn}(t)\delta^{2}(\vec{x})\\
\langle\hat{A}_{i}^{\hat{M}}(t,\vec{x})\hat{A}_{j}^{\hat{N}}(0,\vec{0})\rangle &=-\ii\delta^{\hat{M}\hat{N}}\varepsilon_{ij}\frac{\pi}{k}\text{sgn}(t)\delta^{2}(\vec{x})
\end{align}
\end{subequations}
where, $m,\hat{m},M,\hat{M}$ are color indices\footnote{Recall that the gauge group is $U(N)_k \times U(N)_{-k}$ and the matter fields are in the bifundamental representation. Raised(lowered) $m,n$ indices denote the (anti)fundamental of $U(N)_k$, raised(lowered) $\hat{m},\hat{n}$ indices denote the (anti)fundamental of $U(N)_{-k}$, $M$ is an adjoint index of $U(N)_k$, and $\hat{M}$ is an adjoint index of $U(N)_{-k}$.}, $\varepsilon_{ij}$ is the spatial Levi-Civita symbol such that $\varepsilon_{12}=1$, and $\text{sgn}(t)$ denotes the sign of $t$.

Note that the propagators contain $\delta^2(\vec{x})$ and are therefore ultra-local. For the gauge fields, this is just an artifact of working in temporal gauge. Since Chern-Simons theory is topological, the 2-point functions take the same form in a Carrollian background as they would in Minkowski background so they would not be ultra-local in $R_{\xi}$ gauge, for example. The ultra-local nature of the matter fields, however, is a consequence of the fact that light-cones close up in the Carrollian limit \cite{Bagchi:2022emh}. This leads to a proliferation of delta functions in perturbative calculations which require regularisation \cite{Cotler:2024xhb}. A very natural way to regulate the matter propagators would be to restore the speed of light in the kinetic terms and then take it to zero in the end of the calculation, restoring Carrollian symmetry. In order to preserve gauge invariance, this would require restoring covariant derivatives along spatial directions. We plan to explore this approach more systematically in the future.

\section{Infinite Superconformal Carroll Symmetry in ABJM} \label{abjmsymmetry}

In this section we will present an infinite dimensional extension of the Carrollian superconformal symmetry in \eqref{eq:abjm_carroll_susy_trans} and \eqref{eq:abjm_carroll_susy_conf_trans}. We will refer to this as the extended super-BMS algebra {\footnote{Note that this supersymmetric extension of the BMS or the Conformal Carroll is not unique. We will have more to say about this in the conclusions.}} and verify that it is a symmetry of the Carrollian ABJM theory in \eqref{eq:abjm_carroll}.

\subsection{Extension of Carrollian Supersymmetry}

We can extend the global Carrollian supersymmetry in \eqref{eq:abjm_carroll_susy_trans} to an infinite-dimensional symmetry following the proposal of \cite{Bagchi:2022owq}:
\begin{eqn}\label{eq:abjm_SBMS_G_var}
\delta_G X_A &= \ii z^r\bar{z}^s C^I_{AB}\bar{\psi}^{ B}\epsilon^I\,,\quad
\delta_G X^{ A} = -\ii z^r\bar{z}^s\tilde{C}^{IAB}\bar{\psi}_B\epsilon^I\\
\delta_G \psi_A &= z^r\bar{z}^s C^I_{AB} \gamma^{0}\epsilon^I D_{t}X^{ B}\,,\quad
\delta_G \psi^{ A}=-z^r\bar{z}^s \tilde{C}^{IAB}\gamma^{0}\epsilon^I D_{t}X_B \,, 
\end{eqn}
where we define fermionic generators $G^{I\,a}_{r,s} $ via
\begin{eqn}
[ \bar{G}^{I}_{r,s}\epsilon^I, \Phi] &=  (\text{RHS of \eqref{eq:abjm_SBMS_G_var}} )\,.
\end{eqn}
In a similar manner, the R-symmetry generators $R^{IJ}$ can be defined via
\begin{eqn}
[\frac{1}{2}\alpha^{IJ}R^{IJ},\Phi ] = (\text{RHS of \eqref{eq:abjm_R_sym}}) .
\end{eqn}
The global Carrollian supersymmetry variations in \eqref{eq:abjm_susy_trans_rescaled} are then generated by $Q^{I\,a}=G^{I\,a}_{0,0}$. Moreover it is straightforward to verify that the action in \eqref{eq:abjm_net_rescaled} enjoys the above symmetry:
\begin{eqn}
\frac{2\pi}{k} \delta_{G} S_{\text{Carroll}}&= \int d^3x \text{Tr}\Bigl(D_t \delta_G X^{ A}D_t X_A + D_t X^{ A}D_t \delta_G X_A + \ii \delta_G \psi_{A} (\gamma^0) D_t \psi^{A } + \ii \psi_{A} (\gamma^0)D_t \delta_G \psi^{A } \Bigr)\\
&=\int d^3x \text{Tr}\Bigl(D_t (-\ii \tilde{C}^{IAB} z^r\bar{z}^s\bar{\psi}_B\epsilon^I  )  D_t X_A + D_t X^{ A}D_t (\ii C^I_{AB}z^r\bar{z}^s \bar{\psi}^{B}\epsilon^I ) \\&+ \ii(-z^r\bar{z}^sC^I_{AB}\bar{\epsilon}^I\gamma^0D_tX^B )\gamma^0D_t\psi^A + \ii\bar{\psi}_A \gamma^0D_t (-z^r\bar{z}^s \tilde{C}^{IAB}\gamma^0\epsilon^I D_tX_B) \Bigr)\\
&=\int d^3x \ii\, \partial_t\text{Tr}\Bigl(z^r\bar{z}^s \tilde{C}^{IAB}\bar{\psi}_{A}\epsilon^ID_tX_{B} \Bigr),
\end{eqn}
where we have noted that the gauge fields transform trivally.

Now consider the commutator of two extended supersymmetric variation acting on $X_A$:
\begin{eqn}
[[\bar{G}^I_{r,s} \epsilon^I_1, \bar{G}^J_{r^{\prime},s^{\prime}} \epsilon^J_2],X_A]&=[\bar{G}^I_{r,s} \epsilon^I_1,[\bar{G}^J_{r^{\prime},s^{\prime}} \epsilon^J_2,X_A]]- [\bar{G}^J_{r^{\prime},s^{\prime}} \epsilon^J_2,[\bar{G}^I_{r,s} \epsilon^I_1,X_A]] \\
&= [\bar{G}^I_{r,s} \epsilon^I_1,\ii z^{r^{\prime}}\bar{z}^{s^{\prime}}C^J_{AB} \bar{\epsilon}^J_2 \psi^B] - [\bar{G}^J_{r^{\prime},s^{\prime}}\epsilon^J_2, \ii z^{r}\bar{z}^{s}C^I_{AB} \bar{\epsilon}^I_1 \psi^B] \\
&=-\ii z^{r+r^{\prime}}\bar{z}^{s+s^{\prime}} \Bigl( C^J_{AB} \bar{\epsilon}^J_2(\tilde{C}^{IBC}\gamma^0 \epsilon^I_1 D_t X_C)- C^I_{AB}\bar{\epsilon}^I_1(\tilde{C}^{JBC}\gamma^0 \epsilon^J_2 D_t X_C ) \Bigr)\\
&= \ii  z^{r+r^{\prime}}\bar{z}^{s+s^{\prime}}(C^I_{AB}\tilde{C}^{JBC}+ C^J_{AB}\tilde{C}^{IBC} )  \bar{\epsilon}^I_1 \gamma^0 \epsilon^J_2  D_t X_A \\
&=2\ii z^{r+r\prime}\bar{z}^{s+s^{\prime}} \bar{\epsilon}^I_1 \gamma^0 \epsilon^I_2  D_t X_A.   \\
\end{eqn}
which implies that
\begin{eqn}
[\bar{G}^I_{r,s} \epsilon^I_1, \bar{G}^J_{r^{\prime},s^{\prime}} \epsilon^J_2] &=  2\ii  \bar{\epsilon}^I_1 \gamma^0 \epsilon^I_2 M_{r+r^{\prime},s+s^{\prime}} \\
\end{eqn}
where the derivatives in \eqref{eq:BMS4_diff_op} are promoted to covariant derivatives when acting on fields. 

We will now look at the algebra of the supercharges. It will be convenient to express the result in terms of spinor indices, for which we will use the following conventions following \cite{Benna:2008zy}: un-barred spinors have an upper index labelled by $a,b,c\ldots $ with value $1,2 $, barred spinors have a lower index, and Gamma matrices have index structure $ \gamma^{\mu a}{}_b $. Further, given a spinor $\epsilon^a $, its associated barred spinor defined via $\bar{\epsilon}=\epsilon^{\rm{T}}\gamma^0 $ can be written as $\bar{\epsilon}_a = \varepsilon_{ab}\epsilon^{b} $, where $\varepsilon_{ab} $ is the Levi-Civita symbol with $\varepsilon_{12}=-1 $. Note that $\varepsilon_{ab} $ and its inverse $\varepsilon^{ab} $ can be used be used to lower and raise spinor indices, respectively. Since in our convention, transforming a spinor into its barred form does not involve conjugation, we will denote $\bar{\epsilon}_a=\varepsilon_{ab}\epsilon^b = \epsilon_a $. With these conventions, the fermionic kinetic terms in the relativistic ABJM Lagrangian will then read $\ii \psi_{Aa} \gamma^{\mu}{}^a{}_b D_{\mu} \psi^{Ab} $ \footnote{Note that when we defined $ \bar{\psi}= \psi^{\rm{T}} \gamma^0 $ in previous section, this should be understood as $ \bar{\psi}= \psi^{\rm{T}} \beta $, with $\beta_{ab} $ having a different spinor index structure to $\gamma^0 $, but $\beta=\gamma^0 $ as a numerical matrix equation.}. With these definitions, we may then write commutator of supersymmetric variations as follows:
\begin{eqn}
[\bar{G}^I_{r,s} \epsilon^I_1, \bar{G}^J_{r^{\prime},s^{\prime}} \epsilon^J_2]&= [\epsilon^I_{1\, a} G^{I\,a}_{r,s} ,\epsilon^J_{2\,b} G^{J\,b}_{r^{\prime},s^{\prime}} ] \\
&= -\epsilon^I_{1\, a}\epsilon^J_{2\,b} \{G^{I\,a}_{r,s},G^{J\,b}_{r^{\prime},s^{\prime}} \} \\
&=  2\ii  \epsilon^I_{1\,a} \gamma^0{}^a{}_b \epsilon^{I\,b}_{2} M_{r+r^{\prime},s+s^{\prime}}\\ 
&= -2\ii \epsilon^I_{1\,a} \epsilon^{I}_{2\,b} (\gamma^0){}^{ab} M_{r+r^{\prime},s+s^{\prime}} \\
\end{eqn}
with $(\gamma^{\mu})^{ab}=\varepsilon^{bc} \gamma^{\mu}{}^a{}_c $. Extracting $\epsilon^I_{1\, a}\epsilon^J_{2\,b} $, we conclude that anticommutator of supercharges is
\begin{eqn}\label{eq:SBMS_algebra_no_G_tilde}
\{G^{I\,a}_{r,s},G^{J\,b}_{r^{\prime},s^{\prime}} \} &= 2\ii\delta^{IJ}(\gamma^0)^{ab} M_{r+r^{\prime},s+s^{\prime}}\,. \\
\end{eqn}

\subsection{Extension of Carrollian Superconformal Symmetry}

In a similar manner, we can also extend the superconformal variations in \eqref{eq:abjm_susy_conf_trans_rescaled}: 
\begin{eqn}\label{eq:abjm_SBMS_G_td_var}
\delta_{\tilde{G}} X_A &= \ii z^r\bar{z}^s C^I_{AB}\bar{\psi}^{ B}\gamma^{i}x_{i}\xi^I\,,\quad
\delta_{\tilde{G}} X^{ A} = -\ii z^r\bar{z}^s \tilde{C}^{IAB} \bar{\psi}_B\gamma^{i}x_{i} \xi^I \\
\delta_{\tilde{G}} \psi_A &= z^r\bar{z}^sC^I_{AB} \gamma^{0}(\gamma^{i}x_{i} \xi^I )D_{t}X^{ B} \,,\quad
\delta_{\tilde{G}} \psi^{ A}=-z^r\bar{z}^s \tilde{C}^{IAB}\gamma^{0}\gamma^{i}x_{i} \xi^I D_{t}X_B   \\
\end{eqn}
where we have defined the extended superconformal generator $\tilde{G}^{I\,a}_{r,s} $ via
\begin{eqn}
[\bar{\tilde{G}}^{I}_{r,s} \xi^I, \Phi] &= (\text{RHS of \eqref{eq:abjm_SBMS_G_td_var}} ) \\
\end{eqn}
The global superconformal transformations in \eqref{eq:abjm_susy_conf_trans_rescaled} correspond to $\mcS^{I\,a}=\tilde{G}^{I\,a}_{0,0} $. Notice that the generators in \eqref{eq:abjm_SBMS_G_td_var} and \eqref{eq:abjm_SBMS_G_var} are related via
\begin{eqn}\label{eq:G_G_td_relations}
\tilde{G}_{r,s}^{I}&= -(\gamma_i x^i) G_{r,s}^{I} =-(\gamma_z z+ \gamma_{\bar{z}}\bar{z}) G_{r,s}^{I} = -\gamma_z G_{r+1,s}^{I} -  \gamma_{\bar{z}} G_{r,s+1}^{I},
\end{eqn}
with $ \gamma_{\bar{z}}= \frac{1}{2}\gamma^z = \frac{1}{2}(\gamma^1+\ii \gamma^2) $ and $\gamma_z = \frac{1}{2}\gamma^{\bar{z}} = \frac{1}{2}(\gamma^1-\ii \gamma^2) $. As a result, the extended superconformal generators are not independent of the extended supersymmetry generators.

While supercharges $\tilde{G} $ and $G $ are not independent, it is still worth checking that variations generated by $\tilde{G}$ \eqref{eq:abjm_SBMS_G_td_var} leave the action invariant. To see the invariance of $S_{\text{kin}} $ in \eqref{eq:abjm_kin_rescaled} under the extended superconformal transformation \eqref{eq:abjm_SBMS_G_td_var}, notice that \eqref{eq:abjm_SBMS_G_td_var} can be obtained from \eqref{eq:abjm_SBMS_G_var} by substituting $\epsilon^{I} $ with $x^{i}\gamma_{i}\xi^{I}$ giving
\begin{align}
\frac{2\pi}{k} \delta_{\tilde{G}} S_{\text{kin}}&=\int d^3x \text{Tr}\Bigl(D_t (-\ii \tilde{C}^{IAB} z^r\bar{z}^s\bar{\psi}_B\epsilon^I  )  D_t X_A + D_t X^{ A}D_t (\ii C^I_{AB}z^r\bar{z}^s \bar{\psi}^{B}\epsilon^I ) \\&+ \ii(-z^r\bar{z}^sC^I_{AB}\bar{\epsilon}^I\gamma^0D_tX^B )\gamma^0D_t\psi^A + \ii\bar{\psi}_A \gamma^0D_t (-z^r\bar{z}^s \tilde{C}^{IAB}\gamma^0\epsilon^I D_tX_B) \Bigr)\Big|_{\epsilon^{I}=\gamma_i x^i \xi^{I} } \nonumber \\
&=\int d^3x \ii\, \partial_t\text{Tr}\Bigl(z^r\bar{z}^s \tilde{C}^{IAB}\bar{\psi}_{A}x^i\gamma_i\xi^ID_tX_{B} \Bigr).
\end{align}
Hence, $\tilde{G} $ indeed generates a symmetry of the theory, which extends the global Carroll superconformal symmetry given by \eqref{eq:abjm_carroll_susy_conf_trans}.

\subsection{Extended Super BMS Algebra} 
Let us now derive the algebra of the symmetry transformations in \eqref{eq:abjm_SBMS_G_var} and \eqref{eq:abjm_SBMS_G_td_var}. By repeating the exercise leading to \eqref{eq:SBMS_algebra_no_G_tilde} or using \eqref{eq:G_G_td_relations}, we readily obtain
\begin{subequations}\label{eq:SBMS_algebra}
\begin{align}
& [L_n,G^{I\,a}_{r,s}] = \left[\left(-r+\frac{n+1}{4}\right)\delta^a{}_b-\frac{\ii (n+1)}{2}\mathcal{J}^a{}_b  \right] G^{I\,b}_{r+n,s} \\
& [\bar{L}_n,G^{I\,a}_{r,s}] = \left[\left(-s+\frac{n+1}{4}\right)\delta^a{}_b+\frac{\ii (n+1)}{2}\mathcal{J}^a{}_b  \right]G^{I\,b}_{r,s+n} \\
& [L_n,\tilde{G}^{I\,a}_{r,s}] = \left[\left(-r+\frac{n+1}{4}\right)\delta^a{}_b+\frac{\ii (n+1)}{2}\mathcal{J}^a{}_b\right]  \tilde{G}^{I\,b}_{r+n,s} + \gamma_z {}^a{}_b G^{I\,b}_{r+n+1,s} \\
&[\bar{L}_n,\tilde{G}^{I\,a}_{r,s}] = \left[\left(-s+\frac{n+1}{4}\right)\delta^a{}_b-\frac{\ii (n+1)}{2}\mathcal{J}^a{}_b \right]  \tilde{G}^{I\,b}_{r,s+n} + \gamma_{\bar{z}} {}^a{}_b G^{I\,b}_{r,s+n+1}\\
& \{G^{I\,a}_{r,s},G^{J\,b}_{r^{\prime},s^{\prime}} \} = 2\ii\delta^{IJ}(\gamma^0)^{ab} M_{r+r^{\prime},s+s^{\prime}}, \\
& \{\tilde{G}^{I\,a}_{r,s},\tilde{G}^{J\,b}_{r^{\prime},s^{\prime}} \} = 2\ii \delta^{IJ}(\gamma^0)^{ab} M_{r+r^{\prime}+1,s+s^{\prime}+1} \\
& \{G^{I\,a}_{r,s},\tilde{G}^{J\,b}_{r^{\prime},s^{\prime}} \} = -2\ii \delta^{IJ}((\gamma_z\gamma^0)^{ba}M_{r+r^{\prime}+1,s+s^{\prime}} + (\gamma_{\bar{z}}\gamma^0)^{ba}M_{r+r^{\prime},s+s^{\prime}+1}) \\
&[R^{IJ}, G^{K\,a}_{r,s}]= \delta^{IK}G^{J\,a}_{r,s}-\delta^{JK}\tilde{G}^{I\,a}_{r,s},   \quad
[R^{IJ}, \tilde{G}^{K\,a}_{r,s}]= \delta^{IK}G^{J\,a}_{r,s}-\delta^{JK}\tilde{G}^{I\,a}_{r,s},  \\
&[R^{IJ},R^{KL}] = \delta^{JK} R^{IL} - \delta^{IK} R^{JL} - \delta^{JL} R^{IK} + \delta^{IL} R^{JK}.
\end{align}  
\end{subequations}
where $\mathcal{J}^a {}_b=\frac{1}{2}(\gamma^{12})^a{}_b =-\frac{1}{2}(\gamma^{0})^a{}_b $. Together with $[M,G]=[M,\tilde{G}]=0$ and the bosonic extended BMS algebra given in \eqref{eq:BMS_alg}, this gives the complete $\mathcal{N}=6 $ extended super-BMS algebra. In Appendix \ref{3dalgcontraction}, we perform a Carroll contraction of the $\mathcal{N}=6 $ relativistic superconformal algebra and show how the finite part of the above algebra fits in with the contracted algebra. We encourage the readers to have a look at this. 

Now to some futher details. Note that closure of the above algebra when acting on fields holds on-shell, as discussed in previous subsections. We see that R-symmetry generators do not arise from the commutators of other generators. If we restrict to the global Carrollian superconformal group (giving \eqref{eq:global_carroll_sconf_ferm} and \eqref{Rcomms25}), this can be understood from Carrollian contraction of the 3d superconformal algebra where the R-symmetry generators are not rescaled by $c$, as described in \cite{Bagchi:2022owq, Lipstein:2025jfj} and reviewed in Appendix \ref{3dalgcontraction}. \footnote{If we were to scale the radius of sphere to infinity along with that of AdS, we would have a remenant (non-compact) $R$-symmetry in our fermionic commutators. For more details see  \cite{Bagchi:2022owq}.} 

We can easily verify that the other commutators of the extended super-BMS algebra \eqref{eq:SBMS_algebra} are consistent with the global superconformal Carrollian algebra obtained by contraction of the relativistic algebra (as described in Appendix \ref{3dalgcontraction}) by restricting our attention to the global bosonic generators $L_{\pm 1}, L_0$, their antiholomorphic counterparts $\bar{L}_{\pm 1}, \bar{L}_0$, and the global fermionic modes $G^{I,a}_{0,0}, \tilde{G}^{I\,a}_{0,0}$. Using \eqref{eq:SBMS_algebra}, one immediately finds 
\begin{eqn}
[L_{-1},G^{I\,a}_{0,0}] = 0\, , \quad [L_{-1},\tilde{G}^{I\,a}_{0,0} ] = \gamma_z{}^a{}_b G^{I\,a}_{0,0}, 
\end{eqn}
which reproduces the expected action of translations on supercharges and superconformal charges. Similarly, the action of $L_1$ yields
\begin{eqn}
[L_1,G^{I\,a}_{0,0}] = \gamma_{\bar{z}}{}^a{}_b \tilde{G}^{I\,b}_{0,0}\, , \quad 
[L_1,\tilde{G}^{I\,a}_{0,0}]=0.
\end{eqn}
These relations are precisely the Carrollian analogues of $[K_i, Q^a] = -(\gamma_i)^a {}_b S^b$ and $[K_i,S^a] = 0$ in the relativistic suprconformal algebra (see Appendix \ref{3dalgcontraction} for details). The action of dilatation and rotation generators follows from the commutators with $L_0$ and $\bar{L}_0$. Recalling the identification $D = L_0 + \bar{L}_0$ and $J = i(L_0 - \bar{L}_0)$, one obtains
\begin{eqn}
    [D,{G}^{I\,a}_{0,0}] &= \frac{1}{2}{G}^{I\,a}_{0,0} \,, \quad [D,\tilde{G}^{I\,a}_{0,0}] = -\frac{1}{2}\tilde{G}^{I\,a}_{0,0} \\
    [-\ii J,{G}^{I\,a}_{0,0}] &= -\ii\mathcal{J}^a{}_b {G}^{I\,b}_{0,0}\,,\quad [-\ii J,\tilde{G}^{I\,a}_{0,0}] = -\ii\mathcal{J}^a{}_b \tilde{G}^{I\,b}_{0,0} \\
\end{eqn}
These commutators show that $G^{I,a}_{0,0}$ and $\tilde{G}^{I,a}_{0,0}$ carry dilatation weights $\pm 1/2$, as expected from the superconformal algebra. Moreover, both sets of fermionic generators transform covariantly under spatial rotations, with $\mathcal{J}$ corresponding to the rotation generator in the spin-$1/2$ representation. Further, the action of R-symmetry generator gives,
\begin{eqn}
[R^{IJ}, G^{K\,a}_{0,0}]&= \delta^{IK}G^{J\,a}_{0,0}-\delta^{JK}\tilde{G}^{I\,a}_{0,0} \, , \quad 
[R^{IJ}, \tilde{G}^{K\,a}_{0,0}]= \delta^{IK}G^{J\,a}_{0,0}-\delta^{JK}\tilde{G}^{I\,a}_{0,0} \, , 
\end{eqn}
which exactly recovers \eqref{Rcomms25}. Finally, the algebra of the global fermionic generators are 
\begin{eqn}
\{G^{I\,a}_{0,0},G^{J\,b}_{0,0} \} &= 2\ii\delta^{IJ}(\gamma^0)^{ab} M_{0,0} \\
\{G^{I\,a}_{0,0},\tilde{G}^{J\,b}_{0,0} \} &= -2\ii \delta^{IJ}((\gamma_z\gamma^0)^{ba}M_{1,0} + (\gamma_{\bar{z}}\gamma^0)^{ba}M_{0,1}  )\\
\{\tilde{G}^{I\,a}_{0,0},\tilde{G}^{J\,b}_{0,0} \} &= 2\ii \delta^{IJ}(\gamma^0)^{ab} M_{1,1} \\
\end{eqn}
which exactly recovers \eqref{eq:global_carroll_sconf_ferm} under the identification given in \eqref{eq:BMS4_global_identification}. 
Taken together, the bosonic and fermionic sectors of the global subalgebra precisely match the structure obtained by Carrollian contraction of the relativistic $\mathcal{N} = 6$ superconformal algebra, providing a consistency check of the extended super-BMS algebra.

\section{Discussions and future directions} \label{conclusion}

\subsection*{Summary}
In this paper we analysed the Carrollian limit of a superconformal Chern-Simons matter theory known as the ABJM theory. The motivation for doing so was to deduce a concrete example of flat space holography from AdS$_4$/CFT$_3$. A first step in this direction was taken in \cite{Bagchi:2024efs}, which worked out the Carrollian limit of Chern-Simons theories with scalar matter and showed that they enjoy an infinite-dimensional Carrollian conformal symmetry corresponding to the the extended BMS$_4$ group. The crucial missing ingredient in that work, which we address in this paper, was to include fermions. A priori it is not obvious that taking the speed of light to zero in a relativistic fermion theory will yield a Carrollian theory because relativistic Dirac matrices do not obey the correct algebra on a Carollian spacetime. In fact, we find that there are four different ways of defining the Dirac algebra on a Carrollian manifold due to the degenerate nature of the metric, and one of these possiblities indeed arises from taking the $c\rightarrow0$ limit of relativistic fermions at leading  order (while the others may play a role at subleading order). Moreover we find that in three dimensions, the minimal way to realise this representation of the Carrollian Dirac algebra requires $4\times 4$ matrices. Remarkably, it is possible to recast the $c\rightarrow0$ limit of the ABJM theory in terms of such matrices and we show that the resulting theory has an infinite-dimensional superconformal Carrollian symmetry, providing a supersymmetric generalisation of the toy models constructed in \cite{Bagchi:2024efs}. Given the simplicity of this model, its high degree of symmetry, and its origin in the AdS$_4$/CFT$_3$ correspondence, we consider this to be a promising starting point for realising a dual description of M-theory in flat background. 

\subsection*{Discussion}

We now list some of the pressing questions that follow from our analysis in this current work.

Perhaps the most urgent task is to actually compute an observable in this theory and match it with a bulk calculation in order to check the duality beyond symmetries. In \cite{Lipstein:2025jfj}, it was shown that the Carrollian limit of correlators of protected operators in the ABJM theory correspond to flat space scattering amplitudes in M-theory with four-dimensional kinematics, so this is an obvious target. On the other hand, computing such correlators from first principles is challenging for a number of reasons. First of all, the supersymmetry of the ABJM theory is enhanced non-perturbatively due to monopole operators. It would be interesting to see if such an enhancement persists in the Carrollian limit. Furthermore, perturbative quantisation of Carrollian theories is well-known to encounter singularities requiring regularisation since lightcones close up in the $c\rightarrow0$ limit \cite{deBoer:2023fnj,Cotler:2024xhb,Cotler:2025npu}. One approach to resolve these difficulties may be to expand beyond the electric limit using the speed of light as a regulator. A related question is to explore the relation of the $c\rightarrow 0$ limit to the flat space limit in momentum space \cite{Raju:2012zr,Marotta:2024sce}. We hope to explore this in the future.

Following on from the previous comment, it would also be of interest to explore the fate of Carrollian symmetry at higher orders in the $c$ expansion. As shown in our companion paper \cite{Bagchi:2026lgk}, one can obtain the so-called magnetic theories built out of the upper fermions we encountered in Section \ref{carrollfermi} in the subleading orders of an expansion in powers of $c$. This would lead to more intricate structures in the corresponding supersymmetry and superconformal algebras. e.g. it is expected from the two-dimensional versions \cite{Bagchi:2016yyf, Bagchi:2017cte, Lodato:2016alv} that the supersymmetries and superconformal symmetries will close onto not just the supertranslations that we have seen in our work, but also to superrotations. These have been called magnetic Carroll supersymmetries in the recent literature \cite{Fontanella:2025tbs, Bulunur:2026yav, Ergec:2026baz}. \footnote{Some other recent relevant literature on Super Carrollian or Super BMS symmetries and related theories include \cite{Koutrolikos:2023evq, Kasikci:2023zdn, Zorba:2024jxb, Concha:2024dap, Grumiller:2025rtm, Zheng:2025cuw, Bruce:2026yvw}.} It would be intriguing to figure out whether one can construct variants of ABJM theory with these subleading fermionic contributions. 

Moreover, it would be good to derive the Carrollian superconformal symmetry obtained in this paper from a geometric perspective by solving the (conformal) Killing spinor equations pertaining to Carrollian manifolds and to formulate it in a way that extends the closure of the algebra off-shell. 

Finally, we would like to extend this approach to other canonical examples of the AdS/CFT correspondence, notably AdS$_5$/CFT$_4$ and AdS$_7$/CFT$_6$ \cite{Maldacena:1997re}. The first case relating $\mathcal{N}=4$ SYM to IIB supergravity in AdS$_5 \times$S$^5$ is by far the best understood example of AdS/CFT.  
In contrast, the AdS$_7$/CFT$_6$ correspondence, which relates a certain 6d superconformal theory with $(2,0)$ supersymmetry to M-theory in AdS$_7 \times$S$^4$ is the least understood of these examples since the boundary theory does not have a Lagrangian description, although it is possible to obtain one after dimensional reduction \cite{Lambert:2010iw,Douglas:2010iu,Lambert:2019jwi}. Studying the interplay of dimensional reduction with the Carrollian limit may provide an alternative dual description of M-theory in flat space which is complementary to the one considered here.

\section*{Acknowledgement}

We thank Jan de Boer and Kostas Skenderis for useful discussions. AB was supported partially by a Swarnajayanti Fellowship from Anusandhan National Research Foundation (ANRF) under grant SB/SJF/2019-20/08 and also by an ANRF grant
CRG/2022/006165. AL is supported by the STFC Consolidated Grant ST/X000591/1. SM is supported by Fellowship for Academic and Research Excellence (FARE) at IIT Kanpur.

\newpage

\begin{appendices}

\section{BMS Invariance of Carroll Fermion Action}\label{app:comp}

In this appendix, we will check the invariance of the fermionic term in \eqref{eq:abjm_carroll} under extended BMS symmetry. Using \eqref{eq:BMS_general}, with the rotation generator $J=\frac{1}{2}\tilde{\Gamma}_{12}=\frac{1}{4}[\tilde{\Gamma}_1,\tilde{\Gamma}_2]$ for Carroll spinors, The extended BMS variations of the spinors take the form
\begin{eqn}\label{eq:BMS_carroll_spinor}
[L_n, \Psi]&= \Bigl[z^{n+1}\partial_z +\frac{1}{2}z^n(n+1)(\Delta[\psi]-\frac{\ii}{2}\tilde{\Gamma}_{12} +t\partial_t   )  \Bigr] \Psi  \\
[\bar{L}_n, \Psi]&= \Bigl[\bar{z}^{n+1}\partial_{\bar{z}} +\frac{1}{2}\bar{z}^n(n+1)(\Delta[\psi]+\frac{\ii}{2}\tilde{\Gamma}_{12} +t\partial_t   )  \Bigr] \Psi  \\
[M_{mn},\Psi]&=z^m\bar{z}^n\partial_t \Psi  \\
\end{eqn}
while the extended BMS variations of barred spinors are
\begin{eqn}
[L_n, \bar{\Psi}]&= \Bigl[z^{n+1}\partial_z +\frac{1}{2}z^n(n+1)(\Delta[\psi^*] +t\partial_t   )  \Bigr] \bar{\Psi}  + \frac{\ii}{4}z^n(n+1)\bar{\Psi} \tilde{\Gamma}_{12} \\
[\bar{L}_n, \bar{\Psi}]&= \Bigl[\bar{z}^{n+1}\partial_{\bar{z}} +\frac{1}{2}\bar{z}^n(n+1)(\Delta[\psi^*] +t\partial_t   )  \Bigr] \bar{\Psi} - \frac{\ii}{4}\bar{z}^n(n+1)\bar{\Psi} \tilde{\Gamma}_{12}  \\
[M_{mn},\bar{\Psi}]&=z^m\bar{z}^n\partial_t \bar{\Psi}  \\
\end{eqn}
Using \eqref{3dcspinor}, the transformations reduce to
\begin{eqn}\label{eq:BMS_spinor}
[L_n, \psi] &= \Bigl[z^{n+1}\partial_z  +\frac{1}{2}z^n(n+1)(\Delta[\psi]+t\partial_t -\ii \mathcal{J} ) \Bigr] \psi  \\
[\bar{L}_n, \psi] &= \Bigl[\bar{z}^{n+1}\partial_{\bar{z}}  +\frac{1}{2}\bar{z}^n (n+1) (\Delta[\psi]+t\partial_t + \ii \mathcal{J})  \Bigr]\psi \\
[M_{mn}, \psi] &= z^m\bar{z}^n \partial_t \psi \\
\end{eqn}
with barred spinor (in relativistic spinor basis) transformed via
\begin{eqn}\label{eq:BMS_conj_spinor}
[L_n, \bar{\psi}] &= \Bigl[z^{n+1}\partial_z  +\frac{1}{2}z^n(n+1)(\Delta[\psi^{\dagger}]+t\partial_t) \Bigr] \bar{\psi} +\frac{\ii}{2}z^{n}(n+1) \bar{\psi}\mathcal{J}  \\
[\bar{L}_n, \bar{\psi}] &= \Bigl[\bar{z}^{n+1}\partial_{\bar{z}}  +\frac{1}{2}\bar{z}^n (n+1) (\Delta[\psi^{\dagger}]+t\partial_t)  \Bigr]\bar{\psi} - \frac{\ii}{2}\bar{z}^{n}(n+1) \bar{\psi}\mathcal{J} \\
[M_{mn}, \bar{\psi}] &= z^m\bar{z}^n \partial_t \bar{\psi} \\
\end{eqn}
For the ABJM theory in the Carroll limit, it is sufficient to consider the fermionic kinetic term minimally coupled to a gauge field as given in \eqref{eq:abjm_carroll}
\begin{align}
\begin{split}
  S_{\text{ferm}} &= \int d^3x \bar{\Psi} \tilde{\Gamma}_0 D_t \Psi \\
  &=  \int d^3x \ii \bar{\psi}\gamma^0D_t \psi \\   
\end{split}
\end{align}
with conformal weights $\Delta[\psi]=\Delta[\psi^{\dagger}]=\Delta[\bar{\psi}]=1$, and covariant derivative being simply $D_t\psi = \partial_t\psi +\ii A_t \psi$. To verify the invariance of this action, we also need the BMS variation of the gauge field, which are given in \cite{carroll_csm}. The gauge field transforms as a weight one primary under the generators
\begin{eqn}
[M_{mn},A_t] &= z^m\bar{z}^n \partial_t A_t\\
[L_n,A_t] &=\Bigl[z^{n+1}\partial_z +\frac{1}{2}z^n(n+1)(1+t\partial_t) \Bigr] A_t\\
[\bar{L}_n,A_t] &=\Bigl[\bar{z}^{n+1}\partial_{\bar{z}} +\frac{1}{2}\bar{z}^n(n+1)(1+t\partial_t) \Bigr] A_t\\
\end{eqn}
and those of $A_i $, which are not important for our current discussion as they do not appear in either the action \eqref{eq:ferm_gauge_kin_rescale} or the transformation law of $A_t $. 

\paragraph{Invariance under supertranslations}
Using the above transformation rules, the variation under $M_{mn}$ is given by
\begin{eqn}
\delta_{M_{mn}} S_{\text{ferm}}&= \int d^3x  \Bigl( \ii z^m\bar{z}^n \partial_t \bar{\psi} \gamma^0 D_t \psi + \ii \bar{\psi} \gamma^0 D_t (z^m\bar{z}^n \partial_t \psi) - \bar{\psi} \gamma^0  z^m \bar{z}^n \partial_t A_t \psi \Bigr)\\
&=\int d^3x  \Bigl( \ii z^m\bar{z}^n \partial_t \bar{\psi} \gamma^0 \partial_t \psi + \ii \bar{\psi} \gamma^0 \partial_t (z^m\bar{z}^n \partial_t \psi) \\&- \bar{\psi} \gamma^0  z^m \bar{z}^n \partial_t A_t \psi - \partial_t\bar{\psi} \gamma^0  z^m \bar{z}^n  A_t \psi- \bar{\psi} \gamma^0  z^m \bar{z}^n  A_t \partial_t\psi \Bigr) \\
&=\int d^3x \ii \partial_t\Bigl(z^m \bar{z}^n \bar{\psi}\gamma^0 D_t \psi  \Bigr), \\
\end{eqn}
which is a total time derivative. Hence the action is invariant under supertranslations. 

\paragraph{Invariance under $L_n$}
The variation of action under $L_n $ is given by
\begin{eqn}\label{eq:ferm_Ln_var}
\delta_{L_n}S_{\text{ferm}} &= \int d^3x \ii \Bigl( (z^{n+1}\partial_z \bar{\psi} +\frac{1}{2}z^n(n+1)(1+t\partial_t)\bar{\psi} + \frac{\ii}{2}z^n(n+1)\bar{\psi}\mathcal{J})\gamma^0 D_t \psi \\&+ \bar{\psi}\gamma^0D_t (z^{n+1}\partial_z \psi +\frac{1}{2}z^n(n+1)(1+t\partial_t -\ii \mathcal{J} )\psi) \\&+ \ii \bar{\psi}\gamma^0 (\bar{z}^{n+1}\partial_{\bar{z}}A_t +\frac{1}{2}\bar{z}^n(n+1)(1+t\partial_t)A_t) \psi \Bigr)\\
&=\int d^3x \ii\Bigl(z^{n+1}\partial_z (\bar{\psi}\gamma^0D_t\psi) +\frac{1}{2}z^n(n+1) (2\bar{\psi}\gamma^0D_t\psi +\ii \bar{\psi}\gamma^0 A_t \psi   ) \\& + \frac{1}{2}z^n(n+1)(t\partial_t\bar{\psi}\gamma^0 D_t \psi + \bar{\psi}\gamma^0 D_t (t\partial_t\psi) + \ii \bar{\psi}\gamma^0 t\partial_tA_t \psi ) \\&+\frac{1}{2}z^n(n+1)(\ii \bar{\psi}\mathcal{J}\gamma^0D_t \psi + \bar{\psi}\gamma^0 D_t (-\ii\psi) ) \Bigr)\\
&=\int d^3x\, \ii \Bigl( z^{n+1}\partial_z(\bar{\psi}\gamma^0D_t \psi)+z^n(n+1)(\bar{\psi}\gamma^0D_t \psi) + \frac{1}{2}z^n(n+1)\partial_t(t \bar{\psi}\gamma^0D_t \psi)  \Bigr)\\
&=\int d^3x\, \ii \Bigl( \partial_z(z^{n+1}\bar{\psi}\gamma^0D_t \psi)+ \partial_t(\frac{1}{2}t z^n(n+1) \bar{\psi}\gamma^0D_t \psi)  \Bigr)\\
\end{eqn}
Thus the variation is again a boundary term. The analysis for $\bar{L}_n$ proceeds analogously.

\section{3D Superconformal Algebra and its Carroll contraction} \label{3dalgcontraction}

\subsection*{$\mathcal{N} =1$ Superconformal Algebra}

The three-dimensional $\mathcal{N}=1$ superconformal algebra is the Lie superalgebra $\mathfrak{osp}(1|4)$, whose bosonic subalgebra is the conformal algebra $\mathfrak{so}(3,2)$, the isometry algebra of four-dimensional anti-de-Sitter space. The full superalgebra extends $\mathfrak{so}(3,2)$ by including two real fermionic generators, $\mathcal{Q}_a$ and $\mathcal{S}_a$, which are Majorana spinors of the Lorentz algebra $\mathfrak{so}(2,1)$\footnote{Note that, in 3D, Lorentz group is $SO(2,1) \sim SL(2,\mathbf{R})$. This implies spinor representation is 2-dimensional and real. This also means a Majorana spinor has 2 real components. So when we write $\mathcal{Q}_{a}\,\, (a=1,2)$, we are referring to a real Majorana spinor. There is no need to introduce a separate $\bar{\mathcal{Q}}$ as an independent generator.}. For $\mathcal{N}=1$, there is only one real Majorana supercharge $\mathcal{Q}_a$ per conformal spinor and no internal symmetry is needed to rotate supercherges into each other. So there is no non-trivial R symmetry for $\mathcal{N}=1$. The full set of generators of the three-dimensional $\mathcal{N}=1$ superconformal algebra is
\begin{align}
\{ P_\mu,\, M_{\mu\nu},\, D,\,K_\mu,\, \mathcal{Q}_a,\, \mathcal{S}_a \},
\end{align}
where $P_\mu$ are translations, $M_{\mu\nu}$ Lorentz generators, $K_\mu$ special conformal transformations, $D$ the dilatation, 
and $\mathcal{Q}_a$, $\mathcal{S}_a$ are the Poincar\'e and conformal supercharges\footnote{Also in 3D, the conformal group is locally isomorphic to $Sp(4,\mathbf{R})$. $\mathcal{N}=1$ supersymmetry does have 2 real supercharges $\mathcal{Q}_{a}$ and 2 conformal supercharges $\mathcal{S}_a,$ which combine into a 4-component real spinor of $Sp(4)$.}, respectively. Here greek indices denote spacetime indices and latin indices spinor indices. 

\medskip

\noindent\textit{Bosonic conformal subalgebra:}
\begin{subequations}
\begin{align}
[M_{\mu\nu}, M_{\rho\sigma}] &= 
\eta_{\nu\rho} M_{\mu\sigma} - \eta_{\mu\rho} M_{\nu\sigma}
- \eta_{\nu\sigma} M_{\mu\rho} + \eta_{\mu\sigma} M_{\nu\rho}, \\
[M_{\mu\nu}, P_\rho] &= \eta_{\nu\rho} P_\mu - \eta_{\mu\rho} P_\nu\,,\quad 
[M_{\mu\nu}, K_\rho] = \eta_{\nu\rho} K_\mu - \eta_{\mu\rho} K_\nu, \\
[D, P_\mu] &= P_\mu, \qquad [D, K_\mu] = - K_\mu, \quad 
[P_\mu, K_\nu] = 2(\eta_{\mu\nu} D - M_{\mu\nu}).
\end{align}    
\end{subequations}

\noindent\textit{Lorentz and dilatation action on supercharges:}
\begin{subequations}
    \begin{align}
[M_{\mu\nu}, \mathcal{Q}_a] &= -\tfrac{1}{2} (\gamma_{\mu\nu} \mathcal{Q})_a, 
& [M_{\mu\nu}, \mathcal{S}_a] &= -\tfrac{1}{2} (\gamma_{\mu\nu} \mathcal{S})_a, \\
[D, \mathcal{Q}_a] &= \tfrac{1}{2} \mathcal{Q}_a, 
& [D, \mathcal{S}_a] &= -\tfrac{1}{2} \mathcal{S}_a.
\end{align}
\end{subequations}
This implies $\mathcal{Q}_a$ has conformal weight $+\tfrac{1}{2}$ and $\mathcal{S}_a$ weight $-\tfrac{1}{2}$.

\noindent\textit{Boson--fermion commutators:}
\begin{subequations}
\begin{align}
[P_\mu, \mathcal{S}_a] &=  (\gamma_\mu \mathcal{Q})_a, 
& [K_\mu, \mathcal{Q}_a] &= -(\gamma_\mu \mathcal{S})_a, \\
[P_\mu, \mathcal{Q}_a] &= 0, 
& [K_\mu, \mathcal{S}_a] &= 0.
\end{align}    
\end{subequations}

\noindent\textit{Fermionic anticommutators:}
\begin{subequations}
\begin{align}
\{ \mathcal{Q}_a, \mathcal{Q}_b \} &= 2\ii (\gamma^\mu)_{ab} P_\mu, \\
\{ \mathcal{S}_a, \mathcal{S}_b \} &= 2\ii (\gamma^\mu)_{ab} K_\mu, \\
\{ \mathcal{Q}_a, \mathcal{S}_b \} &= 2\ii \varepsilon_{ab} D 
 + \ii(\gamma^{\mu\nu})_{ab} M_{\mu\nu}.
\end{align}    
\end{subequations}
Here $\gamma_{\mu\nu} = \tfrac{1}{2} [\gamma_\mu, \gamma_\nu]$, and in three dimensions one may use the duality relation
$\gamma_{\mu\nu} = \varepsilon_{\mu\nu\rho} \gamma^\rho$.

\subsection*{Carroll contraction of the 3D $\mathcal N=1$ superconformal algebra}
We perform the Carroll contraction by introducing the speed parameter $c$
via $x^0 = c t$ and taking $c\to 0$ with the following rescalings of generators. Split spacetime indices as $\mu=(0,i)$ with $i=1,2$. Define
\begin{subequations}
\begin{align}
&H = cP_0,\quad P_i = P_i,\quad B_i = cM_{0i} ,\quad J = M_{12} ,\quad
K = cK_0\,,\quad K_i = K_i,\quad D=D.
\end{align}
\end{subequations}

\noindent\textit{Bosonic Carroll–conformal algebra:}
In the limit $c\to 0$, the commutators among the Carroll generators are
\begin{subequations}
\begin{align}
&[J, P_i] = \varepsilon_{ij} P_j, \qquad
[J, B_i] = \varepsilon_{ij} B_j, \qquad
[J, K_i] = \varepsilon_{ij} K_j, \\
&[D, P_i] = P_i, \qquad [D, K_i] = -K_i, \qquad [D, H] = H, \qquad [D, K] = -K,\\
&[B_i, H] = 0, \qquad [B_i, P_j] = -\delta_{ij}H, \qquad [B_i, K_j] = -\delta_{ij} K,\\
&[P_i, K_j] = -2\delta_{ij} D + 2\varepsilon_{ij} J, \qquad [H, K] = -2D,\\
&[P_i,P_j]=0,\qquad [K_i,K_j]=0,\qquad [B_i,B_j]=0,
\end{align}
\end{subequations}
where $\varepsilon_{12}=+1$ and spatial indices are raised/lowered with $\delta_{ij}$.

\medskip

\noindent\textit{Fermionic sector:}
A consistent Carroll contraction requires choosing scalings for $Q_a,S_a$ so the right-hand sides remain finite. A commonly used, finite choice is
\begin{align}
Q_a = {\sqrt{c}}\,\mathcal{Q}_a,\qquad S_a = {\sqrt{c}}\,\mathcal{S}_a,
\end{align}
together with the bosonic scalings above. With these scalings and keeping only finite leading terms as $c\to0$, one obtains the (finite) Carrollian fermionic relations
\begin{subequations}\label{eq:global_carroll_sconf_ferm}
\begin{align}
\{Q_a,Q_b\} &= 2\ii(\gamma^0)_{ab}\,H, \label{qq}\\
\{S_a,S_b\} &= 2\ii(\gamma^0)_{ab}\,K, \label{ss}\\
\{Q_a,S_b\} &= 2\ii(\gamma^{0i})_{ab}\,B_i .\label{qs}
\end{align}
\end{subequations}

Mixed commutators of bosons with fermions are (leading finite terms)
\begin{subequations}
\begin{align}
&[J,Q_a] = -\mathcal{J}_a^{\,\,b}Q_b,\qquad [J,S_a] = -\mathcal{J}_a^{\,\,b}S_b,\\
&[D,Q_a] = \tfrac12 Q_a,\qquad [D,S_a] = -\tfrac12 S_a,\\
&[B_i,Q_a] = [B_i,S_a] = [H,S_a] = [K,Q_a] = 0\\
&[P_i,S_a] = (\gamma_i)_a^{\,\,b}Q_b
,\qquad [K_i,Q_a] = -(\gamma_i)_a^{\,\,b}S_b.
\end{align}
\end{subequations}

\subsection*{Identification with $4$d Poincar\'e Algebra}
We begin with the $\mathcal N=1$ super-Poincaré algebra in four-dimensional Minkowski spacetime, whose generators are
\begin{equation}
\{\,\mathcal P_\mu,\; M_{\mu\nu},\; \mathbb{Q}_\alpha,\; \bar {\mathbb{Q}}_{\dot\alpha}\,\}
\qquad 
\end{equation}
where $\mathcal P_\mu$ generate translations, $M_{\mu\nu}$ Lorentz transformations, and
$\mathbb{Q}_\alpha$, $\bar{\mathbb{Q}}_{\dot\alpha}$ are the Weyl supercharges.

One can identify these generators with those of the three-dimensional
superconformal Carroll algebra. The latter is generated by
\begin{equation}
\{\,H,\; P_i,\; B_i,\; J,\; D,\; K,\; K_i,\; Q_a,\; S_a\,\},
\qquad i=1,2,\quad a=\pm,
\end{equation}
where $H$ is the Carroll Hamiltonian, $P_i$ spatial translations, $B_i$ Carroll boosts,
$J=M_{12}$ rotation, $D$ dilatation, $K,K_i$ special conformal generators, and
$Q_a,S_a$ the supercharges and superconformal charges in the $\{\pm\}$ basis.

\medskip

The explicit linear identification between the four-dimensional
super-Poincaré generators and the three-dimensional superconformal Carroll generators is presented below. 
\begin{subequations}
\begin{align}
D &= M_{03}, 
&\qquad 
J &= M_{12}, 
&\qquad 
B_i &= \sqrt{2}\,\varepsilon_{ij}\,\mathcal P_j,
\\
H &= \sqrt{2}\,(\mathcal P_0 - \mathcal P_3), 
&\qquad 
P_1 &= -\,(M_{02} + M_{23}), 
&\qquad 
P_2 &= \,(M_{01} + M_{13}),
\\
K &= \sqrt{2}\,(\mathcal P_0 + \mathcal P_3), 
&\qquad 
K_1 &= -\,(M_{02} - M_{23}), 
&\qquad 
K_2 &= \,(-M_{01} + M_{13}),
\\
Q_{+} &= \tfrac{1}{\sqrt{2}}\, \mathbb{Q}_1, 
&\qquad 
Q_{-} &= \tfrac{1}{\sqrt{2}}\, \bar{\mathbb{Q}}_1,
\\
S_{+} &= -\tfrac{1}{\sqrt{2}}\, \bar{\mathbb{Q}}_2, 
&\qquad 
S_{-} &= -\tfrac{1}{\sqrt{2}}\, \mathbb{Q}_2.
\end{align}
\end{subequations}

\subsection*{Extension to 3D $\mathcal{N}=6$ Superconformal Algebra}
The three-dimensional $\mathcal{N}=6$ superconformal algebra is the Lie superalgebra $\mathfrak{osp}(6|4)$, whose bosonic subalgebra is the conformal algebra $\mathfrak{so}(3,2)\oplus\mathfrak{so}(6)$. Here $\mathfrak{so}(3,2)$ is the conformal algebra (isometry algebra of $\mathrm{AdS}_4$) and $\mathfrak{so}(6)\cong\mathfrak{su}(4)$ is the R-symmetry algebra that rotates the six supercharges into one another. The full set of generators is 
\begin{align}
\{ P_\mu,\, M_{\mu\nu},\, D,\,K_\mu,\, \mathcal{Q}_a^I,\, \mathcal{S}_a^I,\, R^{IJ} \},
\end{align}
where $P_\mu$ are translations, $M_{\mu\nu}$ Lorentz generators,
$K_\mu$ special conformal transformations, $D$ the dilatation,
$\mathcal{Q}_a^{\,I}$ and $\mathcal{S}_a^{\,I}$ are the Poincar\'e and conformal supercharges carrying a fundamental $\mathbf{6}$ index $I=1,\ldots,6$ of $SO(6)$, and $R^{IJ}=-R^{JI}$ are the 15 generators of the R-symmetry algebra $\mathfrak{so}(6)$. 

The bosonic conformal subalgebra and the action of Lorentz generators and $D$ on the supercharges take the same form as in the $\mathcal{N} =1$ case, now decorated with the index $I$.

\medskip

\noindent\textit{$SO(6)$ R-symmetry algebra:}
\begin{align}
[R^{IJ}, R^{KL}]
  &= \delta^{JK} R^{IL} - \delta^{IK} R^{JL}
   - \delta^{JL} R^{IK} + \delta^{IL} R^{JK}.
\end{align}

\noindent\textit{$SO(6)$ R-symmetry action on supercharges:}
\begin{subequations} \label{Rcomms2}
\begin{align}
[R^{IJ}, \mathcal{Q}_a^{\,K}]
  &= \delta^{IK}\mathcal{Q}_a^{\,J} - \delta^{JK}\mathcal{Q}_a^{\,I}, \\
[R^{IJ}, \mathcal{S}_a^{\,K}]
  &= \delta^{IK}\mathcal{S}_a^{\,J} - \delta^{JK}\mathcal{S}_a^{\,I}.
\end{align}
\end{subequations}
The bosonic generators $P_\mu$, $M_{\mu\nu}$, $D$, $K_\mu$ are all singlets under $R^{IJ}$ and commute with it.

\noindent\textit{Fermionic anticommutators:} The $\{ \mathcal{Q}, \mathcal{S} \}$ anticommutator acquires an $R^{IJ}$ term compared with the $\mathcal{N}=1$ case,
\begin{subequations}
\begin{align}
\{ \mathcal{Q}_a^I, \mathcal{Q}_b^J \} &= 2\ii\delta^{IJ} (\gamma^\mu)_{ab} P_\mu, \\
\{ \mathcal{S}_a^I, \mathcal{S}_b^J \} &= 2\ii\delta^{IJ} (\gamma^\mu)_{ab} K_\mu, \\
\{ \mathcal{Q}_a^I, \mathcal{S}_b^J \} &= 2\ii\delta^{IJ} \varepsilon_{ab} D 
 + \ii\delta^{IJ}(\gamma^{\mu\nu})_{ab} M_{\mu\nu} + 2\ii\epsilon_{ab}R^{IJ},
\end{align}   
\end{subequations}

\subsection*{Carroll contraction of the 3D $\mathcal N=6$ superconformal algebra}
The bosonic and fermionic Carroll scaling is identical to the $\mathcal N=1$ case; they reproduces the supefrconformal algebra unchanged, only now decorated with an index $I$ for $\mathcal{Q}$ and $\mathcal{S}$. A key question is how to treat $R^{IJ}$. For the purpose of this paper, we will assume that R-symmetry generators are not contracted. This, however, is not the only case. From holography perspective, one may wish to attempt to contract the algebra such that five out of $15$ generators become abelian and the remaining $10$ generators rotate supercharges between the $SO(5)$ direction, which forms $ISO(5)$ algebra. We will refer to \cite{Bagchi:2022owq} for detailed discussion on such a non-trivial contraction.

Under the assumption we have, the mixed boson-fermion commutators are the same as in the $\mathcal{N} =1$ case with the addition of the $I$ index. The $R^{IJ}$ part becomes subleading and drops off after contraction, giving
\begin{subequations}\label{eq:global_carroll_sconf_ferm}
\begin{align}
\{Q_a^I,Q_b^J\} &= 2\ii\delta^{IJ}(\gamma^0)_{ab}\,H, \label{qqN6}\\
\{S_a^I,S_b^J\} &= 2\ii\delta^{IJ}(\gamma^0)_{ab}\,K, \\
\{Q_a^I,S_b^J\} &= 2\ii\delta^{IJ}(\gamma^{0i})_{ab}\,B_i.\label{qsN6}
\end{align}
\end{subequations}
R-symmetry actions on the Carrollian supercharges are inherited directly from the relativistic algebra without modification:
\begin{subequations} \label{Rcomms25}
    \begin{align}
    [R^{IJ}, Q_a^K]
  &= \delta^{IK} Q_a^J - \delta^{JK} Q_a^I, \\
[R^{IJ}, S_a^K]
  &= \delta^{IK} S_a^J - \delta^{JK} S_a^I,
    \end{align}
\end{subequations}

\end{appendices}

\bibliographystyle{JHEP}
\bibliography{ref}

\end{document}